\documentclass[sigconf]{acmart}
\usepackage{booktabs} % To thicken table lines
\usepackage[english]{babel}
\usepackage{moresize}
\usepackage{amsmath}
\usepackage{algorithmic}
\usepackage{balance}
\usepackage{comment}
\usepackage{paralist}
\usepackage{bm}
\usepackage{pgfplots}
\usetikzlibrary{pgfplots.dateplot}
\usepackage{flushend}
\usepackage[english]{babel}
\usepackage[latin1]{inputenc}
\usepackage{mathrsfs}
\usepackage{graphicx}

\usepackage{amssymb}
\usepackage{amsfonts}
\usepackage{url}
\usepackage{longtable}
\usepackage{rotating}
\usepackage{multirow}
\usepackage{mathrsfs}
\usepackage{subfigure}
\usepackage{enumitem}
\usepackage[linesnumbered,algoruled,boxed,lined]{algorithm2e}
\usepackage{adjustbox}
\usepackage{hyperref}
\usepackage{pgfplots}
\pgfplotsset{compat=1.18}
\definecolor{tblue}{RGB}{31,119,180}
\definecolor{torange}{RGB}{255,127,14}
\definecolor{tgreen}{RGB}{44,160,44}
\definecolor{tred}{RGB}{214,39,40}
\definecolor{tpurple}{RGB}{148,103,189}

\newcommand{\hide}[1]{} %hide

\newcommand{\ie}{\emph{i.e.,}\xspace}
\newcommand{\eg}{\emph{e.g.,}\xspace}

\newcommand{\trans}{{\mkern-1.5mu\mathsf{T}}}

\newcommand{\model}{KGRec}
\newcommand{\paratitle}[1]{\noindent\textbf{#1}}

\copyrightyear{2023} 
\acmYear{2023} 
\setcopyright{acmlicensed}\acmConference[KDD '23]{Proceedings of the 29th ACM SIGKDD Conference on Knowledge Discovery and Data Mining}{August 6--10, 2023}{Long Beach, CA, USA}
\acmBooktitle{Proceedings of the 29th ACM SIGKDD Conference on Knowledge Discovery and Data Mining (KDD '23), August 6--10, 2023, Long Beach, CA, USA}
\acmPrice{15.00}
\acmDOI{10.1145/3580305.3599400}
\acmISBN{979-8-4007-0103-0/23/08}

\begin{document}


% \fancyhead{}

% \title{Self-Supervised Graph Transformer via Adaptive Masked Autoencoding for Recommendation}

% \title{Adaptive Masked Graph Transformer for Recommendation}

\title[Knowledge Graph Self-Supervised Rationalization for Recommendation]{Knowledge Graph Self-Supervised Rationalization \\for Recommendation}

% \title{Knowledge Graph Self-Supervised Rationalization \\for Recommendation}

\author{Yuhao Yang}
\affiliation{
  \institution{University of Hong Kong}
  % \city{Hong Kong}
  % \country{China}
}
\email{yuhao-yang@outlook.com}

\author{Chao Huang}
\authornote{Chao Huang is the corresponding author.}
\affiliation{
  \institution{University of Hong Kong}
  % \city{Hong Kong}
  % \country{China}
}
\email{chaohuang75@gmail.com}

\author{Lianghao Xia}
\affiliation{
  \institution{University of Hong Kong}
  % \city{Hong Kong}
  % \country{China}
}
\email{aka\_xia@foxmail.com}

\author{Chunzhen Huang}
\affiliation{Wechat, Tencent}
% \affiliation{
%   \institution{Wechat, Tencent}
%   \city{Guangzhou}
%   \country{China}
% }
\email{chunzhuang@tencent.com}


\begin{abstract}

In this paper, we introduce a new self-supervised rationalization method, called \model, for knowledge-aware recommender systems. To effectively identify informative knowledge connections, we propose an attentive knowledge rationalization mechanism that generates rational scores for knowledge triplets. With these scores, \model~integrates generative and contrastive self-supervised tasks for recommendation through rational masking. To highlight rationales in the knowledge graph, we design a novel generative task in the form of masking-reconstructing. By masking important knowledge with high rational scores, \model~is trained to rebuild and highlight useful knowledge connections that serve as rationales. To further rationalize the effect of collaborative interactions on knowledge graph learning, we introduce a contrastive learning task that aligns signals from knowledge and user-item interaction views. To ensure noise-resistant contrasting, potential noisy edges in both graphs judged by the rational scores are masked. Extensive experiments on three real-world datasets demonstrate that \model~outperforms state-of-the-art methods. We also provide the implementation codes for our approach at \url{https://github.com/HKUDS/KGRec}.

\end{abstract}

\maketitle

\section{Introduction}
\label{sec:intro}

% Recommender systems, which connect users to items that they potentially have interests in, have become an influential component in today's Web, considering the information overload problem. And collaborative filtering (CF) which assumes that users with similar interactions share similar interests towards items, is one of the leading paradigms for recommender systems~\cite{vaecf,neumf,bpr,koren2021advances}.

With the rise of information overload, recommender systems have become a critical tool to help users discover relevant items of interest~\cite{bert4rec,yang2023debiased}. Among the leading paradigms in this field is collaborative filtering (CF), which assumes that users with similar interactions share similar interests in items~\cite{vaecf,neumf,xia2023graph}. CF has proven to be effective in a wide range of applications and has driven significant advances in the field of recommender systems.

In recent years, collaborative filtering (CF) frameworks have undergone significant improvements with the introduction of neural networks and latent embedding for users and items, leading to effective enhancements for traditional matrix factorization methods (\eg \cite{neumf, he2017neural, kluver2018rating}). Moreover, novel models that integrate variational autoencoders, attention mechanisms, and graph neural networks have further increased the performance of CF (\eg \cite{vaecf,chen2017attentive,ngcf,he2020lightgcn}). However, the sparsity of user-item interactions fundamentally limits the scope of performance improvement. To address this issue, incorporating a knowledge graph (KG) as a rich information network for items has gained traction in collaborative filtering, leading to knowledge graph-enhanced recommendation.

% The exploration of knowledge-aware recommendation starts with embedding-based methods and path-based methods. To be specific, some studies~\cite{cke,ktup,dkn} inject transition-based knowledge graph embedding (\eg TransR~\cite{transr}) into item embedding to enrich user and item modeling. While some other studies~\cite{per,kprn} focus on extracting semantically meaningful meta-paths from the KG and perform complex modeling of users and items along these meta-paths. In order to unify the embedding-based and path-based methods in a win-win manner, recent researches adopt the powerful graph neural networks (GNNs) to capture multi-hop high-order information by propagation and aggregation on the KG. These state-of-the-art solutions include~\cite{kgat,kgin,kgcn}.

The exploration of knowledge graph-enhanced recommendation begins with embedding-based methods and path-based methods. Specifically, some studies~\cite{cke,ktup,dkn} incorporate transition-based knowledge graph embedding, such as TransR~\cite{transr}, into item embedding to enrich user and item modeling. Other studies~\cite{per,kprn} focus on extracting semantically meaningful meta-paths from the KG and perform complex modeling of users and items along these meta-paths. To unify embedding-based and path-based methods in a mutually beneficial manner, recent research has adopted powerful graph neural networks (GNNs) to capture multi-hop high-order information through propagation and aggregation on the KG. These state-of-the-art solutions include~\cite{kgat,kgin,kgcn}.

\begin{figure}[t]
\centering
\includegraphics[width=\linewidth]{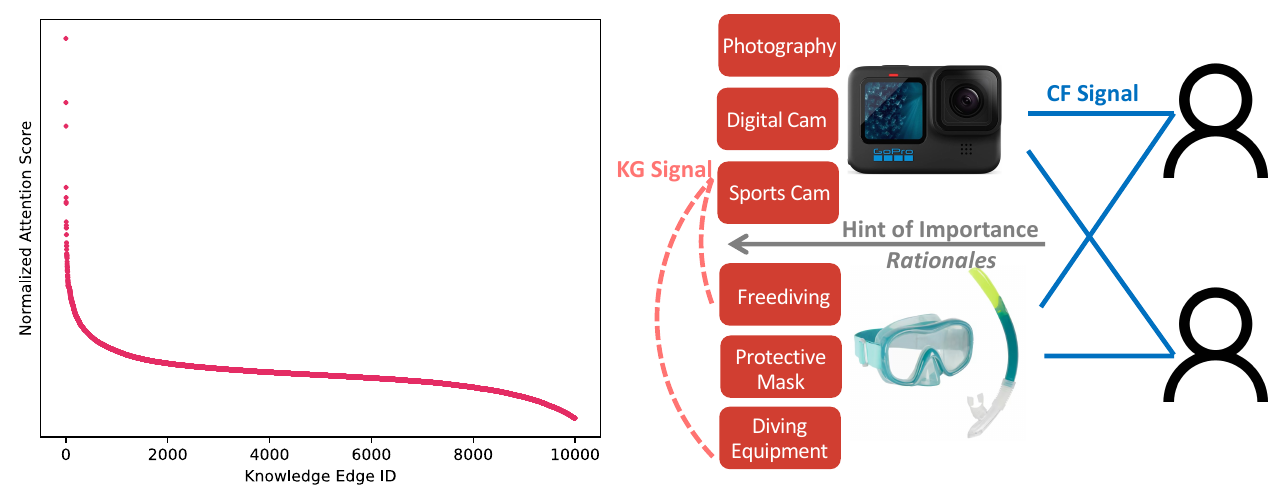}
\vspace{-0.2in}
\caption{The left figure displays a distribution of attentive scores for knowledge triplets in the baseline method of KGAT, which is skewed towards the tail end. On the other hand, the right figure suggests that we can determine the rationality of knowledge triplets for recommendation by analyzing the training labels of user-item interactions.}
\label{fig:intro_case}
\vspace{-0.2in}
\end{figure}

% Despite their effectiveness, the introduction of knowledge graphs brings additional noise and sparsity issues, resulting in further sub-optimal performances~\cite{kopra}. To alleviate the problems, latest studies propose to use contrastive learning (CL) for better knowledge-aware recommendation. For example,~\cite{kgcl} applies stochastic graph augmentation on the KG and performs CL as against noisy entity and long-tail problems in KG.~\cite{kgic,mcclk} design a cross-view CL paradigm between the KG and user-item graph to improve KG representation learning with real labels from recommendation. However, we argue that these methods adopt either simple random augmentation or intuitive cross-view information, failing to notice the important latent rationales between the KG and recommendation task.

Although knowledge graphs have proven effective for improving recommendation systems, they can also introduce noise and sparsity issues, leading to sub-optimal performances~\cite{kopra}. To address these issues, recent studies propose using contrastive learning (CL) for better knowledge-aware recommendation. For example, KGCL~\cite{kgcl} applies stochastic graph augmentation on the KG and performs CL to address noisy entity and long-tail problems in the KG.~\cite{mcclk} design a cross-view CL paradigm between the KG and user-item graph to improve KG representation learning with real labels from recommendation. However, we argue that these methods adopt either simple random augmentation or intuitive cross-view information, failing to consider the important latent rationales between the KG and recommendation task.

% In Figure~\ref{fig:intro_case}, we report distribution of attention scores of knowledge triplets in KGAT on the left, and on the right we give a motivating case that illustrates the rationales in the KG emphasized by CF signals. From the distribution, obviously, only a small proportion of "head" knowledge triplets are highly contributive to recommendation as rationales, by obtaining significant attention scores. Others, however, exhibit a long tail of low scores in the distribution and are thus less informative in the network. We further refer to the case to provide better understanding for hierarchical rationality between KG and CF signals. On an e-commerce platform, users usually buy diving glasses and underwater cameras together. To make correct prediction, connections with common semantics "sports/diving" will be highlighted in the KG. Thus, for the underwater cameras, the knowledge "Photography" and "Digital Cam" will be less important compared to ``Sports Cam''.

Figure~\ref{fig:intro_case} presents the distribution of attention scores of knowledge triplets in KGAT on the left, and a motivating case on the right that illustrates the rationales in the KG emphasized by CF signals. The distribution of attention scores in the KGAT model shows that only a small proportion of knowledge triplets have high attention scores and are thus highly contributive to recommendation as rationales. The remaining knowledge triplets exhibit a long tail of low scores in the distribution and are less informative in the network. To better understand the relationship between KG and CF signals, we provide an example of an e-commerce platform where users often purchase diving glasses and underwater cameras together. To make accurate predictions, the connections with common semantics ``Sports/Diving'' will be highlighted in the KG. Thus, for the underwater cameras, the knowledge ``Photography'' and ``Digital Cam'' will be less important compared to ``Sports Cam''. This highlights the importance of identifying and emphasizing relevant rationales in the KG to improve recommendation performance.

In order to achieve accurate and effective knowledge graph-based recommendations, it is important to explicitly model the rationales behind the user preference learning. To address this challenge, we propose a new knowledge graph-enhanced recommender system, called \model\ to leverage attentive knowledge rationalization to generate task-related rational scores for knowledge triplets. \model\ proposes a self-supervised rationale-aware masking mechanism to extract useful rationales from the KG, by adaptively masking knowledge triplets with higher rational scores. By forcing \model\ to learn to reconstruct these important connections, we highlight task-related knowledge rationales. We also align the rational semantics between the KG signals and the Collaborative Filtering (CF) signals via a knowledge-aware contrasting mechanism. This is achieved by filtering out low-scored knowledge that may be potential noise by masking during graph augmentation for contrastive learning. Finally, we inject the rational scores into the knowledge aggregation for the recommendation task, enabling knowledge rational scores to be learned tightly from the CF labels.

In summary, we make the following contributions in this paper:
\begin{itemize}[leftmargin=*]
    % \item We reveal hierarchical rationality in the knowledge graph for recommendation. On the one hand, useful knowledge triplets account for a small proportion of the whole graph. On the other, knowledge can be rationalized by downstream CF signals.

    \item We unify generative and contrastive self-supervised learning for knowledge graph-enhanced recommender systems, which enables the distillation of the useful knowledge connections within the knowledge graph for recommendation and align them in a noise-free and rationale-aware manner.
    
    % \item We propose a novel model \model~to explicitly model the rationality for knowledge-aware recommendation by unifying generative and contrastive SSL via rational masking. Useful knowledge rationales are highlighted while potential noise is depressed.

    \item Our proposed rationale-aware masking mechanism allows us to identify and highlight the most important and relevant information within the knowledge graph, while suppressing potential noise or irrelevant knowledge graph connections.
    
    % \item We conduct extensive experiments on three real-world datasets to validate the superior design of our \model. We also provide in-depth analysis to justify the reasonableness of our design.

    \item To validate the effectiveness of our proposed model, \model, we conduct extensive experiments on three real-world datasets. Evaluation results provide strong evidence that our proposed model achieves superior performance compared with existing knowledge-aware recommender systems.
    
\end{itemize}
\section{PRELIMINARIES}
We begin by introducing the concepts that will be used in our paper and formally defining the KG-enhanced recommendation task. \\\vspace{-0.12in}

\paratitle{User-Item Interaction Graph.}
In a typical recommendation scenario, we have a set of users, denoted by $\mathcal{U}$, and a set of items, denoted by $\mathcal{V}$. Let $u\in\mathcal{U}$ and $v\in\mathcal{V}$ represent a single user and item, respectively. We construct a binary graph $\mathcal{G}_u = {(u, y_{uv}, v)}$ to denote the collaborative signals between users and items, with $y_{uv}=1$ if user $u$ interacted with item $v$, and vice versa. \\\vspace{-0.12in}

% Consider a typical recommendation scenario, in which a user set $\mathcal{U}$ and item set $\mathcal{V}$ is given. And $u\in\mathcal{U}$ and $v\in\mathcal{V}$ denote single user and item, respectively. We naturally construct a binary graph $\mathcal{G}_u = \{(u, y_{uv}, v)\}$ from the interactions to denote the collaborative signals between users and items, \ie $y_{uv}=1$ if user $u$ interacted with item $v$, and vice versa.

\paratitle{Knowledge Graph.}
We represent real-world knowledge about items with a heterogeneous graph consisting of triplets, denoted by $\mathcal{G}_k = {(h,r,t)}$. $h,t\in\mathcal{E}$ are knowledge entities, and $r\in\mathcal{R}$ represents the semantic relation connecting them, such as (\textit{author, wrote, book}). It is important to note that the item set is a proper subset of the entity set, \ie $\mathcal{V}\subset\mathcal{E}$. This allows us to model the complex relationships between items and entities in the KG. \\\vspace{-0.12in}

% We represent real-world knowledge with a heterogeneous graph with triplets $\mathcal{G}_k = \{(h,r,t)\}$. Particularly, $h,t\in\mathcal{E}$ are knowledge entities and $r\in\mathcal{R}$ represents the semantic relation connecting them, \eg~(\textit{Paris, the capital of, France}). Note that the item set is a proper subset of the entity set: $\mathcal{V}\subset\mathcal{E}$.

\paratitle{Task Formulation.}
Our KG-aware recommendation task can be formally described as follows: given a user-item interaction graph, denoted by $\mathcal{G}_u$, and a knowledge graph, denoted by $\mathcal{G}_k$, our goal is to learn a recommender model, denoted by $\mathcal{F}(u,v|\mathcal{G}_u,\mathcal{G}_k,\Theta)$, where $\mathcal{F}$ represents the model architecture with learnable parameters $\Theta$. The output of the model is a value in the range $[0,1]$ that indicates the likelihood of user $u$ interacting with item $v$.

% For the KG-aware recommendation task, our goal can be formally described as: given a user-item interaction graph $\mathcal{G}_u$ and a knowledge graph $\mathcal{G}_k$, we learn a pairwise rating model as $\mathcal{F}(u,v|\mathcal{G}_u,\mathcal{G}_k,\Theta)$, where $\mathcal{F}$ denotes the model architecture with learnable parameters $\Theta$. The output results in the range $[0,1]$ judges how likely the user $u$ would interact with item $v$.

\section{Methodology}
\label{sec:solution}
\begin{figure*}
    \centering
    \includegraphics[width=\linewidth]{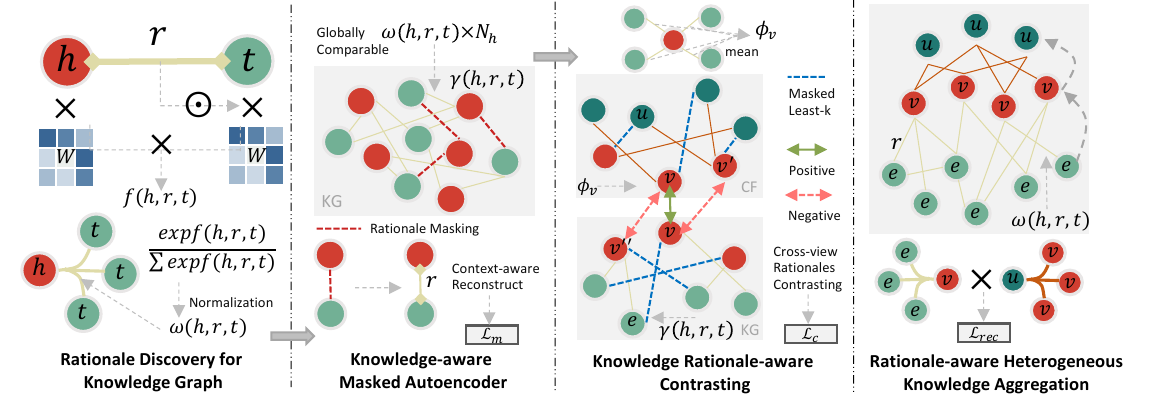}
    \vspace{-0.1in}
    % \caption{The overall framework of \model. Attentive knowledge rationalization generates rational scores for KG triplets. Connections with high rational scores are masked and \model~is trained to reconstruct the important connections under relational context. Low-scored KG triplets are considered as noise to be removed for rationales contrastive learning between u-i and KG graphs. Knowledge aggregation for recommendation is also weighted by the rational scores.}
    \caption{The overall framework of \model. The attentive knowledge rationalization module generates rational scores for KG triplets based on their importance for the recommendation task. Connections with high rational scores are masked, and the model is trained to reconstruct the important connections under relational context. Low-scored KG triplets are considered as noise and removed for rationales contrastive learning between user-item interactions and knowledge graphs.}
    \label{fig:arch}
    \vspace{-0.15in}
\end{figure*}

In this section, we introduce detailed technical design of our proposed~\model. The overall framework is present in Figure~\ref{fig:arch}.

\subsection{Rationale Discovery for Knowledge Graph}
% \subsection{Attentive Knowledge Rationalization}
To automatically distill essential semantics for recommendation from the complex knowledge graph, we propose a rationale weighting function that learns the probability of knowledge triplets being the underlying rationale for collaborative interactions. This rationale function weighs each knowledge triplet based on a learnable graph attention mechanism. Inspired by the heterogeneous graph transformer (HGT)~\cite{hgt}, which discriminates the importance of heterogeneous relations, we implement the rationale weighting function $f(h,r,t)$ as follows:
% To automatically distill important semantics for recommendation from the complex knowledge graph, we device a rationale weighting function that learns the probability of knowledge triplets being the underlying rationale for collaborative interactions. In particular, this rationale function weighs each knowledge triplet based on a learnable graph attention mechanism. Inspired by heterogeneous graph transformer (HGT)~\cite{hgt} that discriminates the importance of heterogeneous edges, given a knowledge triplet $(h,r,t)$, we implement the rationale weighting function $f(h,r,t)$ as follows:

\begin{align}
\label{eq:rational}
    f(h,r,t) =  \frac{\mathbf{e}_h\mathbf{W}^Q \cdot \left(\mathbf{e}_t\mathbf{W}^K \odot \mathbf{e}_r\right)^\trans}{\sqrt{d}}, 
\end{align}
\noindent Here, $\mathbf{e}_h$, $\mathbf{e}_r$, and $\mathbf{e}_t$ are embeddings for the head, relation, and tail entities, respectively. The trainable weights for attention, $\mathbf{W}^Q$ and $\mathbf{W}^K$, have dimensions of $\mathbb{R}^{d\times d}$, where $d$ is the hidden dimensionality. To model the relational context, we use the element-wise product between the relation $r$ and the tail entity $t$, which corresponds to the rotation of the entity embedding $\mathbf{e}_t$ to the latent space of relation $r$~\cite{rotate,kgin}. The rationale score $f(h,r,t)$ of a knowledge triplet indicates its importance in assisting user preference, as learned by the model and guided by the labels from the recommendation task. To ensure comparability of rationale scores across neighbors of the same head entity, we normalize the scores by the number of neighbors $\mathcal{N}_h$ using the following softmax function:

% where $\mathbf{e}_h, \mathbf{e}_r, \mathbf{e}_t$ are embeddings for $h,r,t$ separately. $\mathbf{W}^Q, \mathbf{W}^K \in \mathbb{R}^{d\times d}$ are trainable weights for attention and $d$ is the hidden dimensionality. Note we use the element-wise product between the relation $r$ and tail entity $t$ to model the relational context, which is essentially the rotation of entity embedding $\mathbf{e}_t$ to the latent space of relation $r$~\cite{rotate,kgin}. Comprehensively, as trained by labels from the recommendation task, a knowledge triplet with higher rationale score $f(h,r,t)$ indicates that it is more conducive to making recommendation decisions. To further make rationale scores of knowledge triplets locally comparable across neighbors of the same head entity, we normalize the scores by the number of neighbors $\mathcal{N}_h$ using the following softmax function:
\begin{equation}
\label{eq:kg_w}
    \omega(h,r,t) = \frac{\exp\left(f(h,r,t)\right)}{\sum_{(h,r^\prime,t^\prime)\in \mathcal{N}_h}\exp \left(f\left(h,r^\prime,t^\prime\right)\right)}.
\end{equation}

% \subsection{Rationalized Heterogeneous Knowledge Aggregation}
\subsection{Rationale-aware Heterogeneous\\ Knowledge Aggregation}
\label{sec:kgagg}
A complex KG often contains a large number of real-world knowledge triplets with heterogeneous nature. Inspired by previous works such as~\cite{kgat, kgin, kgcl}, we design an aggregation layer for the knowledge graph that reflects the relational heterogeneity of knowledge triplets. In particular, we focus on the rationales of knowledge triplets, which enable dynamic weighting considering the importance of neighbor entities. To build the knowledge aggregator, we inject the relational context into the embeddings of the neighboring entities, weighting them with the knowledge rationale scores.

% A complex knowledge graph usually contains real-world knowledge triplets with the heterogeneity nature. Inspired by~\cite{kgat, kgin, kgcl}, we design an aggregation layer for the knowledge graph that reflects the relational heterogeneity of knowledge triplets. Rationales of knowledge triplets are another key focus for aggregation, which enables dynamic weighting considering the importance of neighbor entities. In light of the above two points, we build the knowledge aggregator by injecting the relational context into entity neighbor embeddings with the weight of knowledge rationale scores:
\begin{equation}
\label{eq:kgagg}
    \mathbf{e}_h^{(l)} = \frac{1}{|\mathcal{N}_h|} \sum_{(h,r,t)\in \mathcal{N}_h}\omega(h,r,t)\mathbf{e}_r\odot\mathbf{e}_t^{(l-1)},
\end{equation}
\noindent where $l$ denotes the layer of the aggregator, and $\mathcal{N}_h \subseteq \mathcal{G}_k$ is the node-centric sub-graph of first-order neighbors. To inject relational context, we use the same element-wise product as in Equation~\ref{eq:rational} to bridge the gap between aggregation and rationale weighting. By performing such aggregation across the entire knowledge graph, we carefully consider the contextual relationships between knowledge entities and weight neighbor information for the head entity according to normalized rationale scores.

% where $l$ denotes the layer of the aggregator and $\mathcal{N}_h \subseteq \mathcal{G}_k$ is the node-centric sub-graph of first-order neighbors. For the injection of relational context, we keep it consistent with Equation~\ref{eq:rational} as the element-wise product to bridge the gap between aggregation and rationale weighting. By performing such aggregation across the whole knowledge graph, we carefully consider contextual relationships between knowledge entities and weight neighbor information for the head entity according to normalized rationale scores.

% Note that since items are a subset of knowledge entities, we obtain knowledge-aware item representations by aggregating paths $\mathbf{e}_v\leftarrow\mathbf{e}_{t_1}\leftarrow\cdots\leftarrow\mathbf{e}_{t_n}$ on the KG by Equation~\ref{eq:kgagg}. To this end, we consider the role of users on the interaction graph $\mathcal{G}_u$ to model collaborative signals between users and items. Hence, we further generate user embeddings by neighbor aggregation on the user-item interaction graph as follows:

It's worth noting that items are a subset of knowledge entities. Therefore, we obtain knowledge-aware item representations by aggregating paths $\mathbf{e}_v\leftarrow\mathbf{e}_{t_1}\leftarrow\cdots\leftarrow\mathbf{e}_{t_n}$ on the KG using Equation~\ref{eq:kgagg}. To model collaborative signals between users and items, we take into account the role of users in the interaction graph $\mathcal{G}_u$. This allows us to generate user embeddings by aggregating the embeddings of the neighboring items in the user-item interaction graph. Specifically, we use a neighbor aggregation method to obtain the user embedding with the following formulas:

% It's worth noting that items are a subset of knowledge entities. Therefore, we obtain knowledge-aware item representations by aggregating paths $\mathbf{e}v\leftarrow\mathbf{e}{t_1}\leftarrow\cdots\leftarrow\mathbf{e}_{t_n}$ on the KG using Equation~\ref{eq:kgagg}. To model collaborative signals between users and items, we take into account the role of users in the interaction graph $\mathcal{G}_u$.
\begin{equation}
    \mathbf{e}_u^{(l)} = \frac{1}{|\mathcal{N}_u|} \sum_{i\in \mathcal{N}_u}\mathbf{e}_v^{(l-1)},
\end{equation}
\noindent where $\mathbf{e}_u$ and $\mathbf{e}_v$ represent the user embedding and item embedding, respectively. It's important to note that an item $v$ is equivalent to a certain entity $h,t$ in the knowledge graph.

We can define the final representations of users and entities as the summation of aggregated embeddings from different layers:
\begin{equation}
\label{eq:sum}
    \mathbf{e}_h = f_k(\mathcal{G}_k;h) = \sum_l^L\mathbf{e}_h^{(l)};\  \mathbf{e}_u = f_u(\mathcal{G}_u;u) = \sum_l^L\mathbf{e}_u^{(l)},
\end{equation}
\noindent $L$ denotes the number of graph aggregation layers, and $f_*(\cdot , \cdot)$ is the function that generates user or entity representations based on the input graph $\mathcal{G}_u$ or $\mathcal{G}_k$, and certain instances $u$ or $h$.

% \vspace{-0.2in}
\subsection{Knowledge-aware Masked Autoencoder}
\subsubsection{\bf{Rationale Masking Mechanism}}
As related works have revealed~\cite{transa, kbert, shi2018open}, noisy or irrelevant connections between entities in knowledge graphs can lead to suboptimal representation learning. This issue can be particularly problematic in knowledge-aware recommendation systems, where user and item representations are further interrupted by KG noises~\cite{kgcl, kopra}, resulting in inaccurate recommendation results. To eliminate the noise effect in the KG and distill informative signals that benefit the recommendation task, we propose to highlight important knowledge triplets with high rationale scores, as learned in Equation~\ref{eq:rational}.

% As revealed by related works~\cite{transa, kbert, shi2018open}, noisy or irrelevant connections between entities in knowledge graphs can lead to suboptimal representation learning. In knowledge-aware recommendation, the issue can be exacerbated since user and item representations are further interrupted by KG noises~\cite{kgcl, kopra} to derive inaccurate recommendation results. 

% To eliminate the noise effect in KG and distill informative signals that benefit the recommendation task, we propose to highlight important knowledge triplets of high rationale scores learned in Equation~\ref{eq:rational} to guarantee better representation learning. 

Recent studies on masked autoencoders~\cite{he2022masked, feichtenhofer2022masked, xia2023automated} have demonstrated the effectiveness of this approach in enabling models to acquire useful implicit semantics by masking important information during the reconstruction of missing knowledge. Building on these findings, we have designed a generative self-supervised learning task that follows a masking-and-reconstructing approach. During each training step, we mask a batch of knowledge triplets in the KG and reconstruct these relational edges towards a generative self-supervised objective. Additionally, we ensure that the masked triplets have globally high rationale scores, meaning that we mask knowledge that is important for the recommendation task and force the model to learn to reconstruct these connections to highlight useful knowledge for encoding user preference.

% Recent revisiting studies on masked autoencoders (MAEs)~\cite{he2022masked, feichtenhofer2022masked} point out that it allows the model to better acquire useful implicit semantics by masking important information during reconstructing the missing knowledge. Inspired by this, we design a generative self-supervised learning task in the masking-and-reconstructing manner. Specifically, during each training step, we mask a batch of knowledge triplets in the KG, and reconstruct these relational edges towards a generative self-supervised objective. We further require that the masked triplets have globally high rationale scores. In other words, we mask knowledge that is important for the recommendation task, and force the model to learn to reconstruct these connections as to highlight useful knowledge.

To obtain a global measure of the rationale importance of knowledge triplets, we design a criterion. In Equation~\ref{eq:kg_w}, $\omega(h,r,t)$ reflects the local importance of the triplet among all edges to the same head entity $h$. However, the degree of the head entity can influence the value of $\omega$, making it difficult to compare the importance of triplets across the entire KG. To address this issue, we adjust $\omega(h,r,t)$ by multiplying it with the number of head entity neighbors $|\mathcal{N}_h|$. This modification ensures that the importance of the triplet is weighted by the number of connections of the head entity, rather than just its degree. The updated equation is:
\begin{equation}
    \gamma(h,r,t) = |\mathcal{N}_h|\cdot \omega(h,r,t) = \frac{|\mathcal{N}_h|\cdot\exp\left(f(h,r,t)\right)}{\sum_{(h,r^\prime,t^\prime)\in \mathcal{N}_h}\exp \left(f\left(h,r^\prime,t^\prime\right)\right)}.
\end{equation}
\noindent The motivation behind this criterion is to identify the most valuable knowledge triplets across the entire KG. By using the rationale score after softmax, we can determine the relative proportion of a knowledge triplet among its head entity neighbors $\mathcal{N}_h$. We multiply the rationale score with the number of head entity neighbors $|\mathcal{N}_h|$, which makes it globally comparable. By using this approach, we can select the most valuable knowledge triplets across the entire KG based on the value of $\gamma(h,r,t)$. To improve sampling robustness, we add Gumbel noise~\cite{gumbel} to the learned rationale scores.

% The motivation behind is that a knowledge triplet with higher rationale score after softmax would have a higher relative proportion among neighbors of its head entity $\mathcal{N}_h$. And by multiplying the neighbor size we can make the proportion independent of the number of neighbors, thus globally comparable. In this way, we can select the most valuable knowledge triplets across the whole KG by the value of $\gamma(h,r,t)$. We further add Gumbel noise~\cite{gumbel} to the learned rationale scores to improve sampling robustness:
\begin{equation}
    \gamma(h,r,t) = \gamma(h,r,t)-\log\left(-\log(\epsilon)\right);\quad\epsilon\sim\text{Uniform}\left(0,1\right),
\end{equation}
\noindent where $\epsilon$ is a random variable sampled from uniform distribution. Then, we generate a set of masked knowledge triplets by selecting the top $k$-highest rational scores in the KG:
% where $\epsilon$ is a random variable sampled from uniform distribution. Then, we generate a set of masked knowledge triplets by selecting the top $k$-highest rational scores in the KG:
\begin{equation}
\label{eq:mk}
    \mathcal{M}_k = \{(h,r,t)|\gamma(h,r,t)\in\text{topk}(\Gamma;k_m)\},
\end{equation}
where $\Gamma$ represents the distribution of all $\gamma(h,r,t)$. Finally, to create an augmented knowledge graph, denoted by $\mathcal{G}_k^m$, we remove the edges $\mathcal{M}_k$ with low rationale scores from the original knowledge graph $\mathcal{G}_k$. In other words, $\mathcal{G}_k^m$ is obtained by subtracting the set of edges $\mathcal{M}_k$ from the set of edges in $\mathcal{G}_k$, represented by $\mathcal{G}_k \setminus \mathcal{M}_k$.

% $\Gamma$ denotes the distribution of all $\gamma(h,r,t)$. At last, by removing edges $\mathcal{M}_k$ from the knowledge graph $\mathcal{G}_k$, we generate an augmented knowledge graph $\mathcal{G}_k^m$ where $k_m$ triplets with high rationale scores are masked: $\mathcal{G}_k^m = \mathcal{G}_k \setminus \mathcal{M}_k$, where $\setminus$ denotes the set minus.

\subsubsection{\bf{Reconstructing with Relation-aware Objective}}
In order to enable our model to recover crucial knowledge in a self-supervised way, we provide the model with entity embeddings created from the augmented graph $\mathcal{G}_k^m$, and train the model to reconnect the masked knowledge edges. Therefore, we begin by applying rationale-aware knowledge aggregation, as outlined in Equation~\ref{eq:kgagg}, on $\mathcal{G}_k^m$ to produce entity embeddings, in which $k_m$ rationale edges have been removed.
% To facilitate our model with the capacity to recover important knowledge in a self-supervised manner, the model is fed with entity embeddings generated from the augmented graph $\mathcal{G}_k^m$, and is trained to reconnect the masked knowledge edges. Hence, we first apply the rationale-aware knowledge aggregation as in Equation~\ref{eq:kgagg} on $\mathcal{G}_k^m$ to generate entity embeddings with $k_m$ rationale edges removed: 
\begin{equation}
    \mathbf{e}_h = f_k(\mathcal{G}_k^m;h);\ \mathbf{e}_t = f_k(\mathcal{G}_k^m;t),
\end{equation}
% where $f_k(\cdot)$ is the aggregation function on knowledge graph defined in Equation~\ref{eq:sum}. To this stage, knowledge triplets with important rationales $\mathcal{M}_k$ that are invisible during the aggregation can thus serve as self-supervision labels for reconstructing. Considering rich relational heterogeneity in the KG, we force the label triplets to minimize the following dot-product log-loss as reconstructing important rational connections under relational contexts, with $\sigma(\cdot)$ denoting the sigmoid activation function:
The function $f_k(\cdot)$ is the aggregation function on the knowledge graph, as defined in Equation~\ref{eq:sum}. At this point, the knowledge triplets with significant rationale scores, denoted by $\mathcal{M}_k$, which were not visible during the aggregation stage, can be used as self-supervision labels for reconstruction. Given the rich relational heterogeneity in the knowledge graph, we aim to reconstruct the important rational connections under relational contexts. To achieve this, we minimize the following dot-product log-loss for the label triplets, with $\sigma(\cdot)$ representing the sigmoid activation function:
\begin{equation}
\label{eq:mr}
    \mathcal{L}_m = \sum_{(h,r,t)\in \mathcal{M}_k}-\log\left(\sigma\left( \mathbf{e}_h^\trans\cdot\left(\mathbf{e}_t\odot\mathbf{e}_r\right)\right)\right).
\end{equation}

\subsection{Knowledge Rationale-aware Contrasting}
\subsubsection{\bf{Rationale-aware Graph Augmentation}}
% As aforementioned, the hierarchical rationales for knowledge triplets are results of the connection between the KG and user-involved recommendation labels. To further improve the interpretability of the designed knowledge rationalization modules, inspired by~\cite{kgic,mcclk}, we devise to align knowledge graph representations with collaborative filtering signals for explicit modeling of cross-view rationales. To build debiased contrastive views, we first design to recognize and remove weakly task-related edges that are potentially noise in both graphs. 

As explained earlier, the hierarchical rationales for knowledge triplets are derived from the connection between the knowledge graph and user-involved recommendation labels. In order to further enhance the interpretability of the knowledge rationalization modules, we draw inspiration from previous works~\cite{mcclk}. Specifically, we propose to align the representations of the knowledge graph with collaborative filtering signals, which allows us to explicitly model cross-view rationales. To construct debiased contrastive views, we begin by identifying and removing weakly task-related edges that could potentially introduce noise in both graphs.

% For the KG, recall that knowledge triplets with lower rationale scores tend to be less contributive to the recommendation task. Therefore, we augment the KG by removing the noisy triplets:
Regarding the knowledge graph, it is worth noting that knowledge triplets with lower rationale scores tend to have less impact on the recommendation task. Consequently, we aim to improve the quality of the graph by removing the noisy triplets. This augmentation process ensures that the remaining triplets are more informative and have a higher rationale score. By doing so, we can enhance the performance of our model and better capture the underlying relationships between the entities in the graph.
\begin{equation}
    \mathcal{S}_k = \{(h,r,t)|\gamma(h,r,t)\in\text{topk}(-\Gamma;\rho_k)\};\ \mathcal{G}_k^c = \mathcal{G}_k \setminus \mathcal{S}_k,
\end{equation}
% where $\gamma, \Gamma$ are the knowledge attentive scores with Gumbel noise, same as in Equation~\ref{eq:mk}. $-\Gamma$ is the negative $\gamma$ values which turns top-k function to calculate least-k. And $\rho_k$ is the hyperparameter that controls the dropout ratio. $\mathcal{G}_k^c$ is the augmented knowledge graph that is debiased from noise with lower rational scores.
In Equation~\ref{eq:mk}, we introduced the knowledge attentive scores $\gamma$ and $\Gamma$, which are computed with the addition of Gumbel noise. Here, $\Gamma$ represents the distribution of all $\gamma$ values. By taking the negative of $\Gamma$, denoted as $-\Gamma$, we can use the top-k function to calculate the least-k values. The hyperparameter $\rho_k$ controls the dropout ratio during training. We also introduce the augmented knowledge graph $\mathcal{G}_k^c$, which is debiased from noise with lower rationale scores. 

% As for the u-i interaction graph, we aim to remove noisy interactions that are not conducive to cross-view alignment. In particular, we expect to retain interaction edges that clearly reflect the user's interests and thus can better guide knowledge graph rationalization by cross-view contrasting. Considering that the semantics of item embedding can be affected by its linked knowledge in the KG, we devise to weight each interaction edge by considering the rationales of the knowledge triplets connected to the item. Technically, we calculate the mean value of rational scores for knowledge triplets linked with the item to reflect the noise of the interaction edge:
In addition to the knowledge graph, we also aim to improve the quality of the u-i interaction graph by removing noisy interactions that are not conducive to cross-view alignment. Specifically, we want to retain interaction edges that clearly reflect the user's interests and can better guide knowledge graph rationalization through cross-view contrasting. Given that the semantics of item embeddings can be influenced by their linked knowledge in the KG, we propose to weight each interaction edge by considering the rationales of the knowledge triplets connected to the item. This approach allows us to better reflect the noise associated with each interaction edge. To implement this, we calculate the mean value of the rationale scores for all the knowledge triplets linked to the item. This mean value is then used as a weight for the corresponding interaction edge, which helps to distinguish between informative and noisy interactions.
\begin{equation}
    \phi_v = \text{mean}(\{\gamma(h,r,t)|{h=v\vee t=v}\}).
\end{equation}
% Hence, a lower $\phi_v$ indicates that knowledge entities neighboring the item in the KG are relatively less contributive to the recommendation task, causing bias in the item representation. To this stage, we filter our interaction edges with low $\phi_v$ to augment the graph. Here we utilize the multinomial distribution sampling strategy~\cite{huang2022stochastic, li2020federated} to derive more randomized samples for edge dropout to avoid over-fitting on user and item representations. Formally, the process can be defined as:
A lower $\phi_v$ value implies that the knowledge entities neighboring an item in the KG are relatively less contributive to the recommendation task, which can lead to bias in the item representation. To address this issue, we filter our interaction edges using the $\phi_v$ score and augment the graph with only the informative interactions. To avoid overfitting on user and item representations, we adopt a multinomial distribution sampling strategy~\cite{huang2022stochastic, li2020federated} to derive more randomized samples for edge dropout. This approach helps to ensure that the model is not overly reliant on a specific set of interactions and can generalize well to new data. Formally, the process can be defined as follows:
\begin{equation}
    \phi_v^\prime  = \frac{\exp \phi_v}{\sum_v \exp \phi_v};\ \mathcal{S}_u\sim\text{multinomialNR}(\Phi^\prime;\rho_u),
\end{equation}
After calculating the $\phi_v$ score for each item $v$, which represents the mean value of the rationale scores for all the knowledge triplets linked to the item, we apply softmax to obtain a probability distribution $\phi^\prime$ over all items. The resulting distribution $\Phi^\prime$ is used to sample a subset of items without replacement using the multinomial distribution sampling method, denoted as $\text{multinomialNR}(\cdot;\cdot)$. Here, $\rho_u$ denotes the size of the sampled candidates. By following the previous definitions, we can generate the augmented u-i graph as the difference between the original u-i graph $\mathcal{G}_u$ and the set of sampled interactions $\mathcal{S}_u$, \ie $\mathcal{G}_u^c=\mathcal{G}_u\setminus \mathcal{S}_u$.

% where $\phi^\prime$ denotes the probability value of $\phi$ after softmax on all items. $\Phi^\prime$ is the distribution of $\phi^\prime_i$ and $\text{multinomialNR}(\cdot;\cdot)$ is the multinomial distribution without replacement. $\rho_u$ denotes the size of sampled candidates. To this stage, following the previous definition, we can generate the augmented u-i graph as $\mathcal{G}_u^c=\mathcal{G}_u\setminus \mathcal{S}_u$.

\subsubsection{\bf{Contrastive Learning with Cross-View Rationales.}}
% With the augmented knowledge graph and u-i graph, we use different pre-defined aggregators to capture the view-specific node representations for items as the contrastive embeddings. For the u-i interaction view, we utilize the state-of-the-art LightGCN~\cite{he2020lightgcn} module to iteratively capture the high-order information on $\mathcal{G}_u^c$:
With the augmented knowledge graph and u-i graph, we use pre-defined aggregators to capture the view-specific node representations for items as the contrastive embeddings. For the u-i interaction view, we utilize the state-of-the-art LightGCN~\cite{he2020lightgcn} module to iteratively capture high-order information on $\mathcal{G}_u^c$.
\begin{equation}
\mathbf{x}_u^{(l)} = \sum_{v\in\mathcal{N}_u}\frac{\mathbf{x}_v^{(l-1)}}{\sqrt{|\mathcal{N}_u||\mathcal{N}_v|}};\ \mathbf{x}_v^{(l)} = \sum_{u\in\mathcal{N}_v}\frac{\mathbf{x}_u^{(l-1)}}{\sqrt{|\mathcal{N}_u||\mathcal{N}_v|}}.
\end{equation}
% And by summing up the representation from all layers, we obtain the final representation as $\mathbf{x}_v^u = \sum_l^L\mathbf{x}_v^{(l)}$.
% For the augmented knowledge graph, we adopt the rationale-aware knowledge aggregation mechanism as defined in Equation~\ref{eq:sum} to generate the knowledge-view item representations:
We obtain the final contrastive embeddings for items in the u-i view by summing up the representations from all layers of the LightGCN module. For the augmented knowledge graph, we use a rationale-aware knowledge aggregation mechanism to generate the knowledge-view item representations, which take into account the rationale scores associated with the knowledge triplets. 
\begin{equation}
    \mathbf{x}_v^k = f_k(\mathcal{G}_k^c; v).
\end{equation}
% Note that $\mathbf{x}_i^u$ and $\mathbf{x}_i^k$ are from different representation spaces of collaborative signals and knowledge signals, we feed them into two different two-layer MLPs to map them into the same latent space:
It is important to note that the contrastive embeddings $\mathbf{x}_i^u$ and $\mathbf{x}_i^k$ are from different representation spaces, namely the collaborative relational signals and knowledge graph signals. We feed them into two different MLPs to map them into the same latent space.
\begin{equation}
    \mathbf{z}_v^* = \sigma\left(\mathbf{x}_v^{*\trans}\mathbf{W}_1^* + \mathbf{b}_1^*\right)^\trans\mathbf{W}_2^* + \mathbf{b}_2^*,
\end{equation}
where the notation $* \in {u,k}$ denotes view-specific representations, namely $\mathbf{z}_v^u$ and $\mathbf{z}_v^k$. The learnable weights and bias denoted as $\mathbf{W}$ and $\mathbf{b}$. By doing so, we can effectively capture the complementary information from both views.

% Finally, we adopt a contrastive objective to force the alignment of cross-view item representations. To avoid over-fitting and eliminate the false-negative effect, inspired by~\cite{wang2022swift}, we modify the widely used InfoNCE~\cite{infonce} loss to specify one random sample for each view as the negative. Formally, we define our contrastive loss as:
To ensure the alignment of cross-view item representations, we adopt a contrastive objective. To avoid over-fitting and eliminate the false-negative effect, as inspired by~\cite{wang2022swift}, we modify the widely used InfoNCE~\cite{infonce} loss by specifying one random sample for each view as the negative. Formally, we define our contrastive loss as:
\begin{equation}
\label{eq:lc}
    \mathcal{L}_c = \sum_{v\in\mathcal{V}}-\log\frac{\exp(s(\mathbf{z}_v^u, \mathbf{z}_v^k)/\tau)}{\sum_{j\in\{v, v^\prime,v^{\prime\prime}\}}(\exp(s(\mathbf{z}_v^u, \mathbf{z}_v^k)/\tau)+\exp(s(\mathbf{z}_j^u, \mathbf{z}_v^k)/\tau))},
\end{equation}
% where $v^\prime, v^{\prime\prime}$ are stochastically sampled negative candidates for item $v$. $s(\cdot)$ is the similarity measurement which we set to the cosine similarity of normalized vectors. $\tau$ is the temperature hyperparameter that controls the hardness~\cite{khosla2020supervised, sgl} of the contrastive goal.
In the contrastive loss, $v^\prime$ and $v^{\prime\prime}$ are stochastically sampled negative candidates for item $v$. The similarity measurement $s(\cdot)$ is set to the cosine similarity of normalized vectors. The temperature hyperparameter $\tau$ controls the hardness of the contrastive goal~\cite{khosla2020supervised,sgl}.

\subsection{Model Learning and Discussion}
% \subsubsection{\bf{Joint Learning.}}
% For the main recommendation task, the dot product between user and item representations $\hat{y}_{uv} = \mathbf{e}_u^\trans\mathbf{e}_v$ is used as the prediction. In line with baseline models~\cite{kgat,kgcl,kgin}, we adopt the widely used BPR~\cite{bpr} loss to optimize the model parameters as:
For the main recommendation task, we use the dot product between the user and item representations as the prediction, which is denoted as $\hat{y}_{uv} = \mathbf{e}_u^\trans\mathbf{e}_v$. To optimize the model parameters, we adopt the widely used Bayesian Personalized Ranking (BPR) loss to optimize the model parameters as follows:
\begin{equation}
    \mathcal{L}_{rec} = \sum_{(u,v,j)\in\mathcal{D}}-\log\sigma\left(\hat{y}_{uv}-\hat{y}_{uj}\right),
\end{equation}
In the BPR loss, we use the training instances $\mathcal{D} = {(u,v,j)}$, where $v$ is the ground-truth and $j$ is a randomly sampled negative interaction. It is worth noting that we continue to use the entity embeddings $\mathbf{e}_v$ from the masked graph $\mathcal{G}_k^m$ for the recommendation task, rather than performing aggregation on the original knowledge graph again. This is because the masked triplets are generally of small size (\eg 512) compared to the whole graph (\eg millions), and this trick can greatly improve the training efficiency while affecting the representation learning only minimally. Moreover, according to~\cite{li2022scaling, he2022masked}, this setting can increase the difficulty of the main task learning and improve the optimization effect.

% where $\mathcal{D} = \{(u,v,j)\}$ is the training instances with $v$ as the ground-truth and $j$ as the randomly sampled negative interaction. Note that we continue to adopt the entity embeddings $\mathbf{e}_v$ from masked graph $\mathcal{G}_k^m$ for the recommendation task, rather than performing aggregation on the original knowledge graph again. Since the masked triplets are generally of small size (\eg 512) compared to the whole graph (\eg millions), this trick affects little on the representation learning while greatly improves the training efficiency, owing to the effective knowledge rationale discovery. According to~\cite{li2022scaling, he2022masked}, this setting can also increase the difficulty of the main task learning and thereby leads to better optimization effect.

To optimize all three loss functions, we use a joint learning approach with the following overall loss function:
\begin{equation}
    \mathcal{L} = \mathcal{L}_{rec} + \lambda_1\mathcal{L}_m + \lambda_2\mathcal{L}_c,
\end{equation}
where $\lambda_1$ and $\lambda_2$ represent the weight of the mask-and-reconstruction and cross-view contrastive learning tasks, respectively. We omit the notation of L2 regularization terms for brevity.

\subsubsection{\bf{Connection to Alignment and Uniformity.}}
Investigating the alignment and uniformity merits of learned representations by the generative and contrastive tasks is important for providing fundamental support for the proposed~\model~method. Following~\cite{wang2020understanding,yu2022graph}, the mathematical definitions for the uniformity and alignment of learned vector representations are presented below:
% To provide fundamental support for the proposed \model\ method, we investigate the alignment and uniformity merits that \model~brings on learned representations by the generative and contrastive tasks. Following~\cite{wang2020understanding, yu2022graph}, the mathematical definitions for the uniformity and alignment of learned vector representations are as:
\begin{align}
\mathcal{L}_{align} &\coloneqq \mathbb{E}_{(x,y)\sim p^+}[\|\mathbf{x}-\mathbf{y}\|_2^\alpha];\ \alpha>0\label{eq:au1}\\
\mathcal{L}_{uni}
&\coloneqq\log \mathbb{E}_{(x,y)\stackrel{i.i.d.}\sim p}[\exp(-\gamma\|\mathbf{x}-\mathbf{y}\|^2_2)];\ \gamma=1/2\sigma^2,\label{eq:au3}
\end{align}
where the exponent $\alpha$ of the Euclidean distance controls the degree to which the learning algorithm focuses on aligning the learned representations with the positive labels. The distribution $p$ of training data and $p^+$ distribution of positive labels are used to compute the expected value of the alignment loss.

% $\alpha$ is the exponent of the Euclidean distance. $p$ is the distribution of training data and $p^+$ is the distribution of positive labels. 

We first prove that the rational masking-reconstructing task is an explicit alignment for features. According to the generative loss in Equation~\ref{eq:mr}, the optimization of $\mathcal{L}_m$ equals to:
\begin{equation}
    \min\mathbb{E}_{(x,y)\sim p_r^+}\left[\sum_{x,y}\log\left(\sigma\left(\mathbf{x^\trans\mathbf{y}}\right)\right)\right],
\end{equation}
where the variable $\mathbf{x}$ corresponds to the feature vector of the head entity $\mathbf{e}_h$, while the variable $\mathbf{y}$ corresponds to the element-wise product of the feature vectors of the tail entity and relation, given by $\mathbf{e}_r\odot\mathbf{e}_t$. The distribution $p_r^+$ represents the set of knowledge rationales with high rational scores. Note that:
% $\mathbf{x}$ refers to head entity $\mathbf{e}_h$ and y to the element-wise product of tail entity and the relation: $\mathbf{e}_r\odot\mathbf{e}_t$. $p_r^+$ is the distribution of knowledge rationales with high rational scores. Note that:
\begin{equation}
\label{eq:align_trans}
    \|\mathbf{x}-\mathbf{y}\|_2^\alpha = (2-2\cdot\mathbf{x}^\trans\mathbf{y})^\frac{\alpha}{2}.
\end{equation}
Since the generative loss function $\mathcal{L}_m$ is in the form of an alignment loss, as defined in Equation~\ref{eq:au1}, minimizing it leads to the alignment of the masked rationale knowledge triplets.

% Hence, $\mathcal{L}_m$ as the generative loss is in form of alignment as defined in Equation~\ref{eq:au1}. By optimizing $\mathcal{L}_m$, we force masked rationale knowledge triplets to align as reconstruction.

We can further show that the contrastive loss in Equation~\ref{eq:lc} reflects the alignment and uniformity properties. Considering that:
\begin{align}
\label{eq:lc_au}
    \mathcal{L}_c &= \mathbb{E}_{\substack{(x,y)\sim p_c^+ \\ \{x_i^-\}_{i=1}^2\sim p_c}}\left[-\frac{1}{\tau}\mathbf{x}^\trans\mathbf{y} + \log\left(\exp(\mathbf{x}^\trans\mathbf{y}/\tau) + \sum_i\exp(\mathbf{{x_i^-}^\trans\mathbf{x}}/\tau) \right) \right]\\
\label{eq:lc_au_lb}    &\geq \mathbb{E}_{\substack{x\sim p_c \\ \{x_i^-\}_{i=1}^2\sim p_c}}\left[ -\frac{1}{\tau} + \log\left( \exp(\frac{1}{\tau}) + \sum_i\exp(\mathbf{{x_i^-}^\trans\mathbf{x}}/\tau) \right) \right]
\end{align}
The positive pair in Equation~\ref{eq:lc} is denoted as $\mathbf{x},\mathbf{y}$, and the negative samples are denoted as $\mathbf{x^-}$ for brevity. The set of random negative samples ${x_i^-}_{i=1}^2\sim p_c$ in Equation~\ref{eq:lc} is drawn from the distribution $p_c$ of cross-view item representations. As a result, the lower bound of the contrastive loss function $\mathcal{L}_c$ in Equation~\ref{eq:lc_au_lb} is satisfied only if the embeddings $\mathbf{x},\mathbf{y}$ are perfectly aligned, i.e., $\mathbf{x}^\trans\mathbf{y}=1$, which is equivalent to the definition of alignment in Equation~\ref{eq:au1}. If the embeddings satisfy the perfect alignment condition, the optimization of $\mathcal{L}_c$ simplifies to a degenerate form.
% Therefore, the lower bound of $\mathcal{L}_c$ in Equation~\ref{eq:lc_au_lb} is satisfied if and only if the embeddings $\mathbf{x},\mathbf{y}$ is perfectly aligned as $\mathbf{x}^\trans\mathbf{y}=1$, equivalent to the definition of alignment in Equation~\ref{eq:au1}. Suppose that the embeddings satisfy the perfect alignment condition, the optimization of $\mathcal{L}_c$ degenerates to:
\begin{align}
    \min\mathbb{E}_{\substack{x\sim p_c \\ \{x_i^-\}_{i=1}^2\sim p_c}}\left[\log\left(\sum_i\exp(\mathbf{{x_i^-}^\trans\mathbf{x}}/\tau)\right)\right],
\end{align}
% which satisfies the objective in Equation~\ref{eq:au3} as the uniformity loss. Hence, the alignment and uniformity merits in $\mathcal{L}_m$ and $\mathcal{L}_c$ can benefit the representation learning, by forcing the agreement between positive pairs and pushing random instances as negatives. 
The alignment and uniformity properties in the generative loss function $\mathcal{L}_m$ and the contrastive loss function $\mathcal{L}_c$ can benefit representation learning by ensuring that positive pairs are in agreement and that random instances are pushed as negatives. In addition, the proposed knowledge rationalization improves the sampling distribution to be rational and noise-resistant, instead of using a random distribution as in the original forms. By exploiting rationales in the KG, we empower the alignment property with rationality-aware positive pairing ability, which provides better gradients for model learning. Additionally, for cross-view rationales, we remove potential noise to build a noise-free distribution, which eliminates the effect of false negative pairing and improves the contrastive effectiveness. Overall, our \model~is able to derive better alignment and uniformity compared to stochastic methods, which can lead to improved representation for more accurate recommendations.

\section{Evaluation}
\label{sec:eval}
\begin{table}[t]
    \centering
    \caption{Statistics of Three Evaluation Datasets.}
    \vspace{-0.15in}
	\resizebox{\linewidth}{!}{
    \begin{tabular}{lcccc}
    \toprule
    Statistics & Last-FM & MIND & Alibaba-iFashion\\
    \midrule
    \# Users & 23,566 & 100,000 & 114,737\\
    \# Items & 48,123 & 30,577 & 30,040\\
    \# Interactions & 3,034,796 & 2,975,319 & 1,781,093 \\
    \# Density & 2.7e-3 &  9.7e-4 & 5.2e-4\\
    \cmidrule(lr){1-4}
    Knowledge Graph\\
    \cmidrule(lr){1-4}
    \# Entities & 58,266 & 24,733 & 59,156\\
    \# Relations & 9 & 512 & 51\\
    \# Triplets & 464,567 & 148,568 & 279,155\\
    \bottomrule
    \end{tabular}
    }
    \label{tab:dataset}
    % \vspace{-0.2in}
\end{table}
% In this section, in order to evaluate our proposed \model~from multiple perspectives, we conduct extensive experiments corresponding to the following research questions:
In this section, we conduct experiments to answer several research questions related to the proposed \model~framework.
\begin{itemize}[leftmargin=*]
\item\textbf{RQ1}: Can \model~outperform state-of-the-art baseline models of different types in terms of recommendation performance?

\item\textbf{RQ2}: How do the key designs in \model~contribute to its overall performance, and what is its sensitivity to hyperparameters?

\item\textbf{RQ3}: What benefits does \model~bring to tackling task-specific challenges such as cold-start and long-tail item recommendation?

\item\textbf{RQ4}: Can \model~derive interpretability with rational scores?

% , and what is the interpretability of its rational scores?

\end{itemize}

% \begin{itemize}[leftmargin=*]
% \item\textbf{RQ1}: Can \model~obtain superior recommendation performances against state-of-the-art baseline models of different sort?
% \item\textbf{RQ2}: How do key designs of \model~contribute to the overall performances, as well as the sensitivity to the hyperparameters?
% \item\textbf{RQ3}: What benefits does the proposed \model\ framework bring to tackle task-specific key challenges?
% \item\textbf{RQ4}: Can \model~derive explainable results for recommendation?
% \end{itemize}

\subsection{Experimental Setup}
\subsubsection{\bf{Dataset}}
To ensure a diverse and representative evaluation, we use three distinct datasets that reflect real-life scenarios: Last-FM for music recommendations, MIND for news recommendations, and Alibaba-iFashion for shopping recommendations. We preprocess the datasets using the commonly adopted 10-Core approach to filter out users and items with less than 10 occurrences. To construct the knowledge graphs, we employ different methods for each dataset. For Last-FM, we map the items to Freebase entities and extract knowledge triplets, following the techniques used in~\cite{kgat} and~\cite{kb4rec}. For MIND, we collect the knowledge graph from Wikidata\footnote{https://query.wikidata.org/} using the representative entities in the original data, following the approach proposed in~\cite{kopra}. For Alibaba-iFashion, we manually construct the knowledge graph using category information as knowledge, as done in~\cite{kgin}. Table~\ref{tab:dataset} summarizes the statistics of user-item interactions and knowledge graphs for three evaluation datasets. 

% For diverse evaluation, we adopt three datasets that cover different real-life scenarios: Last-FM (music), MIND (news) and Alibaba-iFashion (shopping). We apply 10-Core preprocessing as a common practice to filter out users and items with occurrence count less than 10. To construct knowledge graphs, for Last-FM, following~\cite{kgat,kb4rec}, the items are mapped into Freebase entities to collect knowledge triplets. For MIND, we follow the practice in~\cite{kopra} to collect the knowledge graph on Wikidata\footnote{https://query.wikidata.org/} as given the representative entities in the original data. For Alibaba-iFashion, the knowledge graph is manually constructed by taking the category information as knowledge~\cite{kgin}. We give detailed statistics for the three datasets and their knowledge graphs we use in Table~\ref{tab:dataset}.

\subsubsection{\bf{Evaluation Protocols}}
To ensure fair evaluation, we employ the full-rank setting and divide our dataset into three parts: 70\% for training, 10\% for hyperparameter tuning, and 20\% for testing. We measure the performance of our proposed \model~using the Recall@N and NDCG@N metrics, with N set to 20 for top-N recommendations. We implement \model~using PyTorch and compare its performance with various baseline models using official or third-party code. To optimize the performance of \model, we conduct a hyperparameter search for the masking size, keeping proportion for contrastive learning, and temperature value. Specifically, we explore values of masking size from the range of $\{128,256,512,1024\}$, keeping proportion $\rho_k$ and $\rho_u$ from $\{0.4,0.5,0.6,0.7,0.8\}$, and temperature value from the range of $\{0.1,\cdots,1.0\}$. The number of GNN layers is set to 2 for all graph-based methods.

% To avoid bias in metrics from negative sampling in evaluation~\cite{krichene2022sampled}, we report performance metrics under the full-rank setting as in consistent with previous works~\cite{kgat,kgin,kgcl}. To ensure the validity of the results, we split the dataset by the proportion of 70\%, 10\% and 20\% for training, hyperparameter tuning and testing, separately. We adpot two test-of-time metrics for top-N recommendation, Recall@N and NDCG@N. Performances are reported at N=20, which is a widely-used value~\cite{neumf,ngcf,kgat}. We implement our proposed \model~in the PyTorch environment. For the compared baseline models, we refer to either their official code release, or third-party re-implementation such as RecBole~\cite{zhao2021recbole}. \model~and all baseline models share common basic hyperparameter settings, such that the embedding size is fixed at 64 and learning rate at $1e^{-3}$. For models that integrate GNNs, the context hop is set to 2 as the same with KGIN, MCCLK, and \model, considering the lower efficiency of heterogeneous graph aggregation. For \model, specifically, we search the masking size $k_m$ for the generative learning in the set of $\{128,256,512,1024\}$. The keeping (not masked) proportion for contrastive learning $\rho_k$ and $\rho_u$ are searched amongst $\{0.4,0.5,0.6,0.7,0.8\}$. The temperature value $\tau$ is searched from the range $\{0.1,\cdots,1.0\}$.

\subsubsection{\bf{Baseline Models}}
% We conduct benchmark evalution between \model~and diverse baseline models in different research lines to verify the effectiveness of our design.

To verify the effectiveness of our proposed design, we conduct benchmark evaluations between \model~and various baseline models from different research lines.

\paratitle{General Collaborative Filtering Methods.}
\begin{itemize}[leftmargin=*]
    \item  \textbf{BPR}~\cite{bpr} is a matrix factorization method that uses pairwise ranking loss based on implicit feedback.
    \item \textbf{NeuMF}~\cite{neumf} incorporates MLP into matrix factorization to learn the enriched user and item feature interactions.
    \item \textbf{GC-MC}~\cite{gcmc} considers recommendation as a link prediction problem on the user-item graph and proposes a graph auto-encoder framework for matrix completion.
    \item \textbf{LightGCN}~\cite{he2020lightgcn} is a state-of-the-art recommendation method based on graph neural networks (GNNs), which improves performance by removing activation and feature transformation.
    % \item \textbf{SGL}~\cite{sgl} introduces the self-supervised learning paradigm to GNN-based recommendation by stochastic augmentation on the user-item graph and InfoNCE self-contrastive objective.
    \textbf{SGL}~\cite{sgl} introduces a self-supervised learning paradigm to GNN-based recommendation by using stochastic augmentation on the user-item graph based on the InfoNCE objective. 
\end{itemize}

\paratitle{Embedding-based Knowledge-aware Recommenders.}
\begin{itemize}[leftmargin=*]
    % \item\textbf{CKE}~\cite{cke} is an embedding-based KG recommender which enriches item representations by TransR~\cite{transr} training on structural knowledge to empower collaborative filtering.
    \item \textbf{CKE}~\cite{cke} is an embedding-based KG recommender that leverages TransR~\cite{transr} to enrich item representations by training on structural knowledge, thereby enhancing collaborative filtering.
    % \item\textbf{KTUP}~\cite{ktup} trains TransH~\cite{transh} task with preference-injected CF, and enables mutual complementation between CF and KG signals.
    \item \textbf{KTUP}~\cite{ktup} trains TransH~\cite{transh} using preference-injected CF and enables mutual complementation between CF and KG signals.
\end{itemize}

\paratitle{GNN-based Knowledge Graph-enhanced Recommenders.}
\begin{itemize}[leftmargin=*]
    % \item\textbf{KGNN-LS}~\cite{kgnnls} considers user preference towards different knowledge triplets in graph convolution, and further introduces label-smoothing as regularization to force similar user preference weights between near items in the KG.
    \item \textbf{KGNN-LS}~\cite{kgnnls} considers user preferences towards different knowledge triplets in graph convolution and introduces label smoothing as regularization to force similar user preference weights between nearby items in the KG.
    % \item\textbf{KGCN}~\cite{kgcn} aggregates knowledge for item representations by considering high-order information with GNN, and it considers preferences from user embedding as weights.
    \item \textbf{KGCN}~\cite{kgcn} aggregates knowledge for item representations by considering high-order information with GNN and uses preferences from user embeddings as weights.
    % \item\textbf{KGAT}~\cite{kgat} introduces the idea of collaborative knowledge graph to apply attentive aggregation on the user-item-entity joint graph. The attention scores reflect the importance of knowledge triplets.
    \item \textbf{KGAT}~\cite{kgat} introduces the concept of a collaborative KG to apply attentive aggregation on the joint user-item-entity graph, with attention scores reflecting the importance of knowledge triplets.
    % \item\textbf{KGIN}~\cite{kgin} is a state-of-the-art method that models user intents for relations and designs relational path-aware aggregation to effectively capture rich information on CKG.
    \item \textbf{KGIN}~\cite{kgin} is a state-of-the-art method that models user intents for relations and employs relational path-aware aggregation to effectively capture rich information on the knowledge graph.
\end{itemize}

\paratitle{Self-Supervised Knowledge-aware Recommenders.}
\begin{itemize}[leftmargin=*]
    % \item\textbf{MCCLK}~\cite{mcclk} is a recently proposed method that performs contrastive learning in a hierarchical way to mine useful structural for user-item-entity graph and its subgraphs.
    \item \textbf{MCCLK}~\cite{mcclk} performs contrastive learning in a hierarchical manner for data augmentation, so as to consider structural information for the user-item-entity graph.
    % \item\textbf{KGCL}~\cite{kgcl} introduces graph contrasitve learning for KG to depress potential knowledge noise. KG contrastive signals are further used to denoise the user-item graph.
    \item \textbf{KGCL}~\cite{kgcl} introduces graph contrastive learning for KGs to reduce potential knowledge noise. KG contrastive signals are further used to guide the user preference learning.
\end{itemize}

% Note that since KGIC~\cite{kgic} is a contemporaneous work comparable both in methodology and performances to MCCLK, we only include the latter for performance benchmark.

\subsection{RQ1: Overall Performance Comparison}
\begin{table}[]
	\centering
	\caption{The overall performance evaluation results for \model~and compared baseline models on three experimented datasets, where the best and second-best performances are denoted in bold and borderline, respectively.}
	\vspace{-0.05in}
	\label{tab:results}
	\resizebox{\linewidth}{!}{
	\begin{tabular}{ccccccc}
		\toprule
		\multirow{2}{*}{Model} & \multicolumn{2}{c}{Last-FM} & \multicolumn{2}{c}{MIND} & \multicolumn{2}{c}{Alibaba-iFashion} \\
        \cmidrule(lr){2-3}\cmidrule(lr){4-5}\cmidrule(lr){6-7}
        ~ & Recall & NDCG & Recall & NDCG & Recall & NDCG \\
		\midrule\midrule
		BPR & 0.0690 & 0.0585 & 0.0384 & 0.0253 & 0.0822 & 0.0501 \\ 
		NeuMF & 0.0699 & 0.0615 & 0.0308 & 0.0237 & 0.0506 & 0.0276 \\ 
		GC-MC & 0.0709 & 0.0631 & 0.0386 & 0.0261 & 0.0845 & 0.0502 \\
		LightGCN & 0.0738 & 0.0647 & 0.0419 & 0.0253 & 0.1058 & 0.0652 \\
		SGL & 0.0879 & 0.0775 & \underline{0.0429} & 0.0275 & 0.1141 & 0.0713 \\
\midrule		
  CKE & 0.0845 & 0.0718 & 0.0387 & 0.0247 & 0.0835 & 0.0512 \\
        KTUP & 0.0865 & 0.0671 & 0.0362 & 0.0302 & 0.0976 & 0.0634\\
\midrule		
  KGNN-LS & 0.0881 & 0.0690 & 0.0395 & \underline{0.0302} & 0.0983 & 0.0633 \\
		KGCN & 0.0879 & 0.0694 & 0.0396 & 0.0302 & 0.0983 & 0.0633 \\
		KGAT & 0.0870 & 0.0743 & 0.0340 & 0.0287 & 0.0957 & 0.0577 \\
		KGIN & 0.0900 & \underline{0.0779} & 0.0357 & 0.0225 & 0.1144 & \underline{0.0723} \\
  \midrule
		MCCLK & 0.0671 & 0.0603 & 0.0327 & 0.0194 & 0.1089 & 0.0707\\
        KGCL & \underline{0.0905} & 0.0769 & 0.0399 & 0.0247 & \underline{0.1146} & 0.0719 \\
		\midrule
		{\model} & \textbf{0.0943} & \textbf{0.0810} & ~\textbf{0.0439} & \textbf{0.0319} & \textbf{0.1188} & \textbf{0.0743} \\
		\bottomrule
	\end{tabular}
	}
	\vspace{-0.05in}
\end{table}

% We report the overall performance benchmark for all methods on the three datasets in Table~\ref{tab:results}. We summarize the following observations based on the results:

We report the performance of all the methods on three datasets in Table~\ref{tab:results}. Based on the results, we make the following observations:

\begin{itemize}[leftmargin=*]

    % \item The proposed \model~consistently outperforms all baseline models on both metrics and all three datasets. We attribute the result to: i) by rational masking and reconstruction, \model~is able to capture knowledge information that is truly useful to the recommendation task. ii) \model~is equipped with rational cross-view contrastive learning on augmented noise-free graphs. Latent relatedness between the KG and CF signals can be better exploited. And iii), the knowledge aggregation layer is weighted by knowledge rational scores to be reflective of different importance of knowledge triplets. Additionally, given that the adopted datasets differ significantly in statistics, the superior results suggest that the proposed knowledge rationalization mechanism can automatically discover useful knowledge related to downstream tasks, regardless of the data characteristics.

    \item The proposed \model~consistently outperforms all baseline models on both metrics and all three datasets. This can be attributed to three factors. First, by using rational masking and reconstruction, \model~is able to capture knowledge information that is truly useful for the recommendation task. Second, \model~is equipped with rational cross-view contrastive learning on augmented, noise-free graphs, which allows for better exploitation of the latent relatedness between KG and CF signals. Third, the knowledge aggregation layer is weighted by knowledge rational scores to reflect the different importance of knowledge triplets. Additionally, the superior results on datasets with vastly different statistics suggest that the proposed knowledge rationalization mechanism can automatically discover useful knowledge related to downstream tasks, regardless of the data characteristics. \\\vspace{-0.12in}
    
    % \item On the three datasets and two metrics, there is no consistent winner among baseline models. Contrastive learning-based methods (\ie MCCLK and KGCL) are not always better than non-self-supervised methods (\eg KGIN). This might be due to the limitation of random graph augmentation or intuitive handcrafted cross-view pairing, which may fail to discover truly useful information from the contrastive views.

    \item On the three datasets, there is no consistent winner among the baseline models. Contrastive learning-based methods (\eg MCCLK and KGCL) are not always better than non-self-supervised methods (\eg KGIN). This may be due to the limitations of random graph augmentation or intuitive handcrafted cross-view pairing, which may fail to discover truly useful KG information from the contrastive views for encoding the interests of users. \\\vspace{-0.12in}
    
    % \item GNN-based knowledge-aware recommenders can consistently beat embedding-based models. This advantage originates from GNNs' ability to capture more complex and higher-order information on the KG, compared to the linear transition-based modeling adopted by embedding-based models.

    \item GNN-based knowledge-aware recommenders can consistently outperform embedding-based models. This advantage is due to GNNs' ability to capture more complex and higher-order information on the KG, compared to the linear transition-based modeling adopted by embedding-based models. \\\vspace{-0.12in} 
    
    % \item The introduction of knowledge graphs does not guarantee better performances. For example, the performances of CKE and KTUP are generally worse than non-KG methods LightGCN and SGL. Even KGNN-LS and KGCN can not in some metrics outperform SGL. This effect becomes more pronounced when the dataset has a complex KG and sparse interactions. We suggest that this is because some KG-aware recommenders fail to effectively model complex relational paths and alleviate noise in the KG. And suboptimal KG representation learning leads to worse recommendation performances. On the contrary, LightGCN and SGL focus more on solving sparsity problem in the original interactions, thus outperform those methods.

    \item The introduction of knowledge graphs does not always lead to better performance in recommendation systems. For instance, methods such as CKE and KTUP typically perform worse than non-KG methods like LightGCN and SGL. Even KGNN-LS and KGCN cannot consistently outperform SGL in some metrics. This effect is more noticeable when the dataset has a complex KG and sparse interactions. We suggest that some KG-aware recommenders struggle to effectively model complex relational paths and mitigate noise in the KG, resulting in suboptimal KG representation learning and worse performances. On the other hand, LightGCN and SGL focus more on resolving the sparsity problem of user-item interactions with self-supervision signals. 
    
\end{itemize}

\subsection{RQ2: Ablation Study}
\subsubsection{\bf{Key Module Ablation}}
\begin{table}[]
    \centering
    \caption{Ablation results of \model~with different variants. The superscript $^\ast$ denotes the largest change in performance.}
    \vspace{-0.1in}
	\resizebox{\linewidth}{!}{
    \begin{tabular}{lcccccc}
    \toprule
    \multirow{2}{*}{Ablation Settings} & \multicolumn{2}{c}{Last-FM} & \multicolumn{2}{c}{MIND} & \multicolumn{2}{c}{Alibaba-iFashion}\\
    \cmidrule(lr){2-3}\cmidrule(lr){4-5}\cmidrule(lr){6-7}
    ~ & Recall & NDCG & Recall & NDCG & Recall & NDCG\\
    \midrule
    \model & \textbf{0.0943} & \textbf{0.0810} & \textbf{0.0439} & \textbf{0.0319} & \textbf{0.1188} & \textbf{0.0743}\\
    \cmidrule(lr){1-7}
    w/o MAE & 0.0918$^\ast$ & 0.0792$^\ast$ & 0.0374$^\ast$ & 0.0238$^\ast$ & 0.1178$^\ast$ & 0.0737$^\ast$\\
    w/o Rationale-M & 0.0929 & 0.0805 & 0.0423 & 0.0311 & 0.1183 & 0.0739\\
    w/o CL & 0.0926 & 0.0796 & 0.0425 & 0.0313 & 0.1180 & 0.0734\\
    w/o Rationale-Aug & 0.0931 & 0.0801 & 0.0405 & 0.0278 & 0.1185 & 0.0741\\
    \bottomrule
    \end{tabular}
    }
    \label{tab:ablation}
    \vspace{-0.2in}
\end{table}

% In this part, we investigate the effectiveness of key modules in the proposed \model~from the aspects of masked autoencoding and contrastive learning. We build four model variants to compare with the original method:

In this study, we investigate the effectiveness of key modules in our proposed \model~from the perspectives of our designed rational masked autoencoding and contrastive learning for recommendation. To compare with the original method, we built four model variants, including:

\begin{itemize}[leftmargin=*]
    \item w/o MAE: removing the generative SSL task of rationale-aware knowledge graph masking and reconstruction.
    \item w/o Rationale-M: replacing the rationale knowledge masking with random masking while keeping the masking size unchanged.
    \item w/o CL: disabling the cross-view contrastive learning task.
    \item w/o Rationale-Aug: replacing the rational graph augmentation with random masking while keeping the masking size unchanged.
\end{itemize}

We report the results of the ablation study in Table~\ref{tab:ablation} and make the following observations: i) The proposed rationale knowledge masking and reconstruction contributes the most to performance enhancement. This demonstrates that mask\&reconstruction is an effective strategy for exploiting highly useful knowledge triplets for recommendation. ii) The rational masking mechanism for both reconstruction and contrastive learning can further improve performance by selecting valuable information and dropping informative knowledge. iii) The contrastive learning is also beneficial for performance. However, we observed that adding non-rationale augmented graph contrastive learning on the MIND dataset can hurt performance. This indicates that simple intuitive cross-view contrasting is not always effective due to noises in the graph.

% iii) The contrastive learning is also positive to the performance. Note that we observe that adding non-rationale augmented graph contrastive learning on MIND dataset can hurt the performance. This indicates that simple intuitive cross-view contrasting is not always effective, due to noises in the graph.

\subsubsection{\bf{Sensitivity to Key Hyperparameters}}
We present our results and discussion of parameter study in Appendix~\ref{sec:a:hp}.

\subsection{RQ3: Model Benefits Investigation}
\subsubsection{\bf{Cold-start Recommendation}}
\begin{figure}[t]
\centering
\subfigure[Alibaba-iFashion@Recall]{
\label{fig:cold_user:recall}
\includegraphics[width=0.48\linewidth]{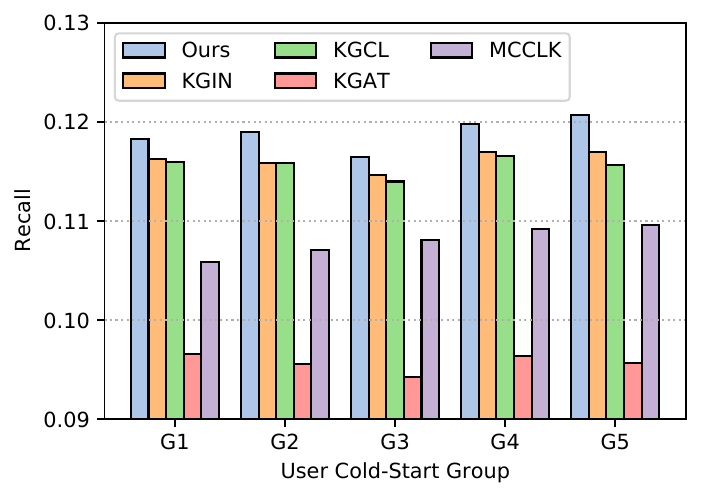}}
\subfigure[Alibaba-iFashion@NDCG]{
\label{fig:cold_user:ndcg}
\includegraphics[width=0.48\linewidth]{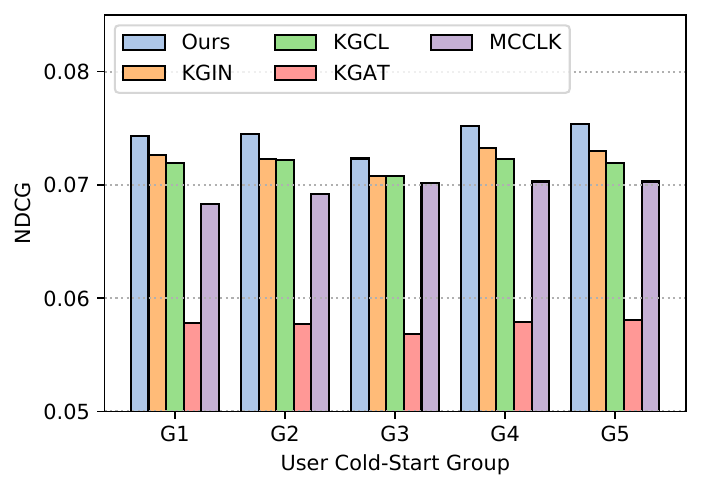}}
\vspace{-0.15in}
\caption{Evaluation results on different user groups. Lower group number implies stronger cold-start effect.}
\label{fig:cold_user}
\vspace{-0.15in}
\end{figure}

% Here we first study how effective is \model~against the common user cold-start problem in recommendation. To be specific, we divide users in Alibaba-iFashion dataset into five groups according to the number of interactions, where smaller group number indicates stronger cold-start effect of users. In different groups, we test performances of \model\ and some strongest baselines separately, and report the results in Figure~\ref{fig:cold_user}. From the results, we observe that \model\ can benefit recommendation for users from all different cold-start groups, comparing to other baseline methods. The reason behind this can be attributed to our design of rationale knowledge masked autoencoding, and the rationale-based cross-view contrastive learning. By highlighting useful knowledge for representation learning and contrasting between cross-view signals, \model~can better alleviate the cold-start and popularity-bias issues for different user groups.

We conduct a study to evaluate the effectiveness of \model~in addressing the common cold-start problem in recommendation systems. We divided users in the Alibaba-iFashion dataset into five groups based on the number of interactions, with smaller group numbers indicating stronger cold-start effects. We then separately tested the performance of \model~and several strong baselines in each group and reported the results in Figure~\ref{fig:cold_user}. Our findings demonstrate that \model~outperforms other baseline methods in all cold-start groups, indicating its effectiveness in addressing the cold-start problem for a diverse range of users. This can be attributed to the design of the rationale knowledge masked autoencoding and rationale-based cross-view contrastive learning, which highlight useful knowledge for representation learning and contrast cross-view signals. Therefore, \model~can effectively alleviate cold-start issue.

\subsubsection{\bf{Long-tail Item Recommendation}}
\begin{figure}[t]
\centering
\subfigure[Alibaba-iFashion@Recall]{
\label{fig:cold_item:recall}
\includegraphics[width=0.48\linewidth]{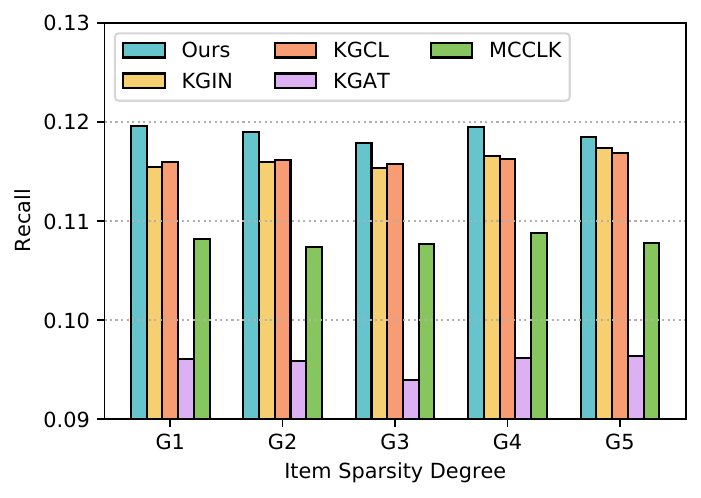}}
\subfigure[Alibaba-iFashion@NDCG]{
\label{fig:cold_item:ndcg}
\includegraphics[width=0.48\linewidth]{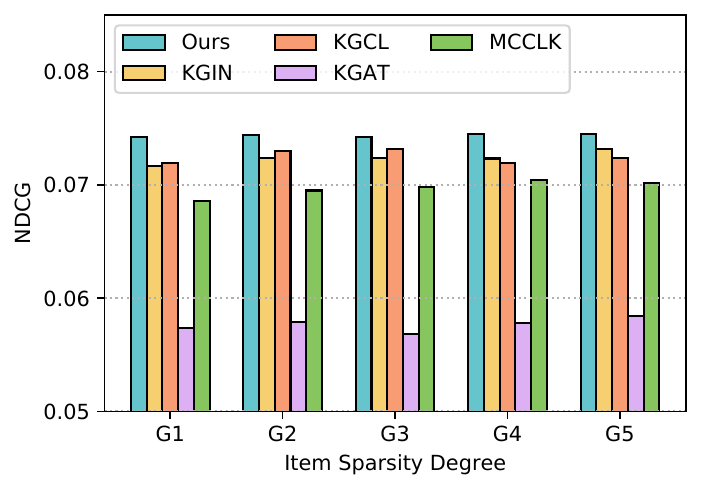}}
\vspace{-0.15in}
\caption{Evaluation results on different item sparsity levels.}
\label{fig:cold_item}
\vspace{-0.15in}
\end{figure}

% We further investigate on whether \model~can benefit the representation learning for long-tail items. We first count the occurrence for each items, and similarly divide all users into five groups according to the average sparsity degree of items that they interacted with. The results are reported in Figure~\ref{fig:cold_item}. From the results, we can observe that \model\ still gains performance improvement consistently across different groups, compared with baseline models. This further illustrates the superiority of \model\ design against data scarcity problems by making better use of external knowledge with rationale mining.

We investigate whether \model~can improve representation learning for long-tail items. We counted the occurrence of each item and divided all users into five groups based on the average sparsity degree of items they interacted with. The results are reported in Figure~\ref{fig:cold_item}. Our findings demonstrate that \model~consistently outperforms baseline models across different groups, indicating its effectiveness in addressing data scarcity problems. This can be attributed to the design of rationale mining, which allows \model~to better leverage external knowledge and improve representation learning for long-tail items.

\subsubsection{\bf{Recommendation with small proportion of KG}}
\begin{figure}[t]
\centering
\subfigure[Last-FM]{
\label{fig:kg_less:lfm}
\includegraphics[width=0.48\linewidth]{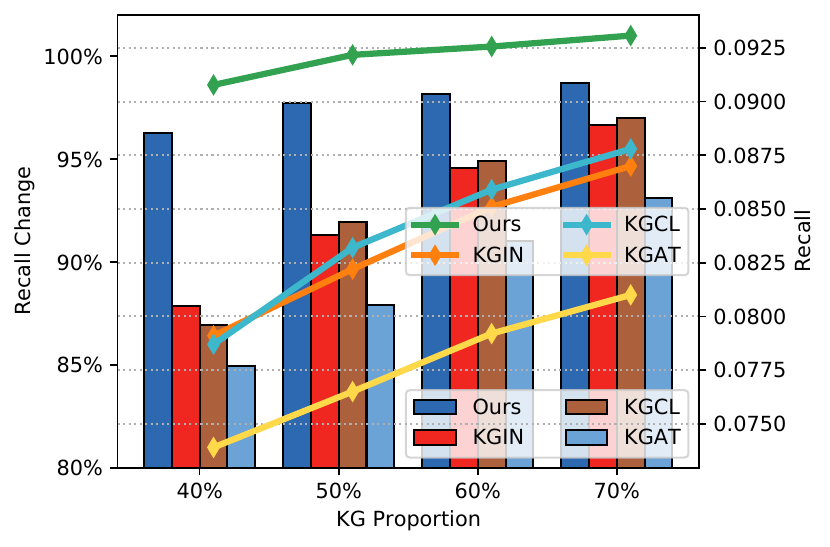}}
\subfigure[Alibaba-iFashion]{
\label{fig:kg_less:ali}
\includegraphics[width=0.48\linewidth]{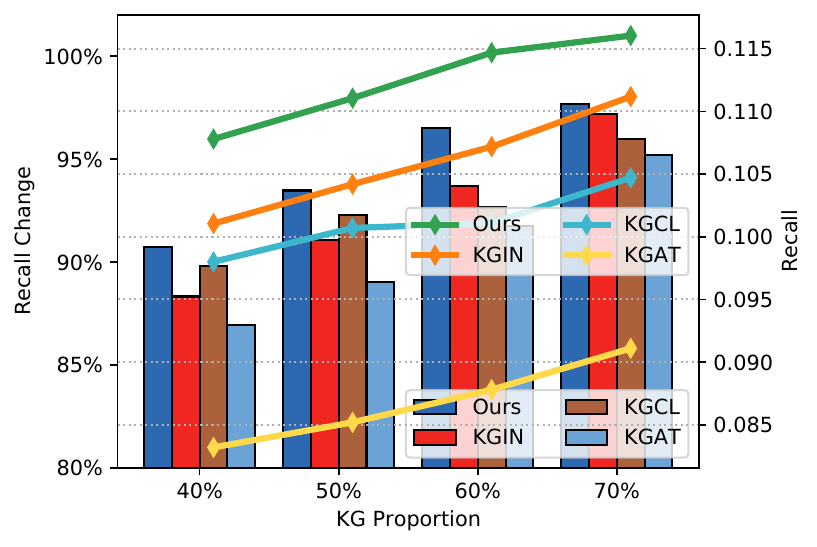}}
\vspace{-0.15in}
\caption{Evaluation results on different KG proportions.}
\label{fig:kg_less}
\vspace{-0.15in}
\end{figure}

% To further investigate the capacity of \model~in highlighting important task-related connections from the knowledge graph, we test recommendation performances for \model~and baseline models under partial knowledge graphs with different keeping ratios ranging in [40\%,50\%,60\%,70\%]. Specifically, we randomly select a proportion of knowledge triplets from the original KG in Last-FM and Alibaba-iFashion datasets for knowledge aggregation. The results are reported in Figure~\ref{fig:kg_less}. We can find that \model~can still maintain considerable performance (>95\% on Last-FM and >90\% on Alibaba-iFashion) with only a small portion of KG. Compared to baseline models, \model~shows minimal performance degradation in all cases. We assume the reason is that rationale knowledge masking and reconstruction mechanism can effectively distill useful knowledge from the given portion of the KG. Also, the cross-view contrastive learning can enhance KG learning with CF signals to alleviate the absence of some knowledge. Overall, the results further verify the rationality of the \model~design.

We evaluate \model's capacity in highlighting important task-related connections from the knowledge graph. Specifically, we tested the recommendation performance of \model~and baseline models under partial knowledge graphs with different keeping ratios ranging from 40\% to 70\%. We randomly selected a proportion of knowledge triplets from the original KG in the Last-FM and Alibaba-iFashion datasets for knowledge aggregation, and the results are reported in Figure~\ref{fig:kg_less}. Our findings demonstrate that \model~can still maintain considerable performance (>95\% on Last-FM and >90\% on Alibaba-iFashion) with only a small portion of KG. Compared to baseline models, \model~shows minimal performance degradation in all cases. This can be attributed to the design of rationale knowledge masking and reconstruction mechanism, which can effectively distill useful knowledge from the given portion of the KG. Additionally, the cross-view contrastive learning can enhance KG learning with CF signals to alleviate the absence of some knowledge. Overall, the results further validate the rationality of \model's design.

% \subsubsection{Alignment and Uniformity Visualization}
% \label{exp:au}
\subsection{RQ4: Model Explainability Study}
\begin{table}[]
    \centering
    \caption{KG Relations with highest average global rationale scores for news categories learned on MIND dataset.}
    \vspace{-0.15in}
    \begin{tabular}{cccc}
    \toprule
Category & Relation (Wiki ID) & Avg. Score \\
\hline
\multirow{2}{*}{\textit{sports}} & member of sports team (P54) & 1.235\\
~ & league of (P118) & 1.117\\
\cmidrule(lr){1-3}
\multirow{2}{*}{\textit{newspolitics}} & member of political party (P102) & 1.341 \\
~&position held (P39) & 1.097\\
\cmidrule(lr){1-3}
\multirow{2}{*}{\textit{travel}} & part of (P361) & 1.105 \\
~& located in (P131) & 1.190\\
\cmidrule(lr){1-3}
\multirow{2}{*}{\textit{finance}} & owned of (P1830) & 1.203 \\
~& stock exchange (P414) & 1.157\\
\cmidrule(lr){1-3}
\multirow{2}{*}{\textit{tv-celebrity}} & award received (P166) & 1.084 \\
~& cast member (P161) & 1.139\\
\bottomrule
\end{tabular}
\label{tab:casestudy}
\vspace{-0.15in}
\end{table}

We discuss the explainability of results generated by \model~in Appendix~\ref{sec:a:case_study}, which provides insights into how \model~incorporates the KG and rationale knowledge for enhancing recommendation.

% We discuss the explainability of recommendation results generated by \model~in Appendix~\ref{sec:a:case_study}.
\section{Related Work}
\label{sec:relate}
\subsection{Knowledge-aware Recommender Systems}

% Knowledge graphs are useful side information for item representation learning and user modeling in recommender systems. Current knowledge-aware recommendation methods can be categorized into three groups: embedding-based methods, path-based methods and GNN-based methods. Embedding-based methods~\cite{cke,dkn,ktup,mkr} incorporate knowledge graph entity embedding into user and item representations as to enhance the recommendation learning. As an example, the representative work CKE~\cite{cke} proposes to integrate the modeling of different types of side information for items with collaborative filtering. It encodes a knowledge graph with the transitive KG completion method TransR~\cite{transr} as part of item representations. Path-based methods~\cite{per,kprn,mcrec} focus on exploiting the rich semantics in relational meth-paths on the KG. For example, KPRN~\cite{kprn} adopts a LSTM to model the extracted meta-paths, and aggregates user preference along each path by fully-connected layers. Path-based models can leverage more complex information along paths in the KG, but intuitively designed meta-paths limits the generalizability of the model, and aggregating along multiple paths on a large graph can be very time-consuming.

Knowledge graphs are valuable sources of side information for item representation learning and user modeling in recommender systems. Currently, knowledge-aware recommendation methods can be generally categorized into three groups: embedding-based methods, path-based methods, and GNN-based methods. i) Embedding-based methods~\cite{cke,dkn,ktup,mkr} incorporate knowledge graph entity embedding into user and item representations to enhance the recommendation learning. For example, CKE~\cite{cke} proposes to integrate the modeling of different types of side information for items with collaborative filtering. It encodes a knowledge graph with the transitive KG completion method TransR~\cite{transr} as part of item representations. ii) Path-based methods~\cite{per,kprn,mcrec} focus on exploiting the rich semantics in relational meta-paths on the KG. For instance, KPRN~\cite{kprn} adopts an LSTM to model the extracted meta-paths and aggregates user preference along each path by fully-connected layers. iii) GNN-based methods~\cite{kgat,kgcn} extend GNNs to model the KG and use the learned representations for recommendation. For example, KGAT~\cite{kgat} proposes to use a graph attention mechanism to propagate user and item embeddings on the KG, and then apply a multi-layer perceptron to produce the final recommendation score.

% To unify the two paradigms and fuse their merits, the research line of GNN-based knowledge-aware recommenders~\cite{kgnnls,kgcn,kgat,kgin} takes advantage of GNNs' powerful ability of capturing high-order information to effectively extract useful information from the KG. KGCN~\cite{kgcn} samples a fixed number of neighbors as the receptive field to aggregate item representations on the KG. The following approach KGAT~\cite{kgat} leverages the idea of graph attention networks (GATs) to weight the knowledge aggregation on the KG by considering differed importance of knowledge neighbors. KGIN~\cite{kgin} further considers user latents towards different relations in the KG and inject relational embedding in the aggregation layer to improve the performance. Overall, GNN-based methods are current state-of-the-art solutions due to their ability in exploiting rich semantics from the graph and considerable efficiency.

The line of GNN-based knowledge-aware recommenders~\cite{kgnnls,kgcn,kgat,kgin} aims to unify the two paradigms and combine their strengths. GNNs have a powerful ability to capture high-order information, making them effective at extracting useful information from the KG. KGCN~\cite{kgcn} samples a fixed number of neighbors as the receptive field to aggregate item representations on the KG. KGAT~\cite{kgat} leverages graph attention networks (GATs) to weight the knowledge aggregation on the KG by considering the different importance of knowledge neighbors. KGIN~\cite{kgin} further considers user latents towards different relations in the KG and injects relational embedding in the aggregation layer to improve performance. GNN-based methods are currently the state-of-the-art solutions due to their ability to exploit rich semantics from the graph and their considerable efficiency.

% \vspace{-0.1in}
\subsection{Self-Supervised Recommendation}

Incorporating self-supervised learning (SSL) techniques into recommender systems has become a new trend in the research community to address inherent data sparsity problems by leveraging additional supervision signals from raw data~\cite{s3rec,li2023graph}. Existing studies have explored various SSL techniques for different recommendation tasks. For large-scale industry applications, \cite{googlecl} introduces contrastive learning in the two-tower architecture for feature augmentation with the proposed correlated feature masking strategy. SGL~\cite{sgl} applies graph contrastive learning to graph collaborative filtering using random augmentation on graphs such as node dropout, edge dropout, and random walk to generate contrastive views and enforce agreement with InfoNCE loss. For sequential recommendation, S3Rec~\cite{s3rec} aims to augment the sequence itself by masking and adopts the contrast between augmented sequences as an auxiliary task. For social recommendation, MHCN~\cite{yu2021self} performs contrastive learning between user embedding and its social embedding extracted from a sub-hypergraph of the social network. For multi-modal recommender systems, MMSSL~\cite{wei2023multi} aims to provide a universal solution for capturing both modality-specific collaborative effects and cross-modality interaction dependencies, allowing for more accurate recommendations.

KGCL~\cite{kgcl} develops graph contrastive learning on the KG to alleviate noise and long-tail problems, while also leveraging additional signals from KG agreement to guide user/item representation learning. MCCLK~\cite{mcclk} employ cross-view contrastive learning between the KG and interaction graph to mitigate sparse supervision signals. However, we argue that these methods do not sufficiently consider the rationales embedded in the KG. By explicitly rationalizing knowledge triplets for recommendation, our \model~achieves a significant performance improvement compared to these methods.
\section{Conclusion}
\label{sec:conclusoin}
% In this paper, we present a novel knowledge graph self-supervised rationalization method (\model) for knowledge-aware recommendation. The motivation lies in the hierarchical rationality of knowledge triplets. We build our method on the attentive knowledge rationalization to weight knowledge triplets, and introduce a novel rational masking and reconstruction module to emphasize rational knowledge. The rational scores are further used to facilitate the cross-view contrastive learning, where low-scored less informative knowledge is filtered out as noise. Results of extensive experiments validate the advantages of \model~against state-of-the-art solutions. In future works, we will explore more complex methods for knowledge graph rationalization, such as graph structure learning and graph sparsification. Another problem worth exploring is how to distill knowledge for effective knowlegde-aware recommendation.

In this paper, we presented a novel graph self-supervised rationalization method (\model) for knowledge-aware recommendation. Our motivation is rooted in the hierarchical rationality of knowledge triplets. We build our method on the attentive knowledge rationalization to weight knowledge triplets, and introduce a novel rational masking and reconstruction module to emphasize rational knowledge. The rational scores were further used to facilitate the knowledge-aware cross-view contrastive learning, where low-scored less informative knowledge was filtered out as noise. Results of extensive experiments validate the advantages of \model~against state-of-the-art solutions. In future works, we will explore more complex methods for knowledge graph rationalization, such as graph structure learning and graph sparsification. This direction can potentially provide more insights into the underlying knowledge graph structure.

\clearpage
\bibliographystyle{ACM-Reference-Format}
\balance
\bibliography{sample-base}

\clearpage
\appendix \section{Appendix}
% \balance
\label{sec:appendix}

\subsection{Sensitivity to Key Hyperparameters}
\label{sec:a:hp}
\begin{figure}[h]
% \vspace{-0.2in}
\centering
\subfigure[Masking size $k_m$]{
\label{fig:hp:msize}
\includegraphics[width=0.31\linewidth]{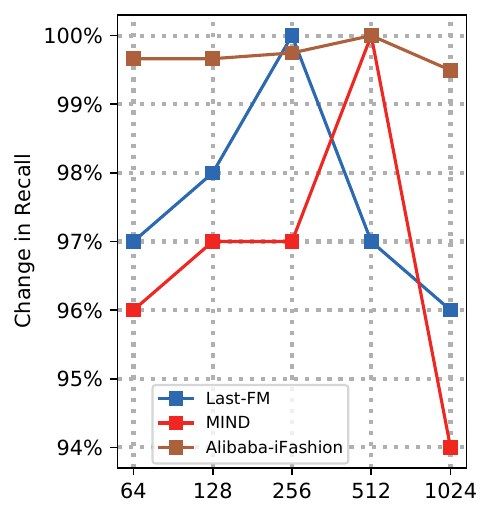}}
\subfigure[CL keep ratio $\rho$]{
\label{fig:hp:rho}
\includegraphics[width=0.31\linewidth]{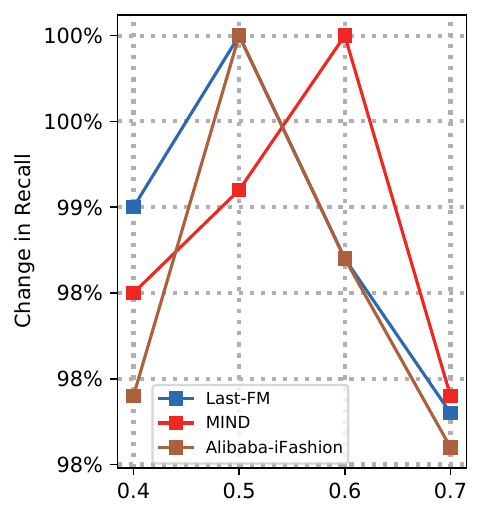}}
\subfigure[CL temperature $\tau$]{
\label{fig:hp:tau}
\includegraphics[width=0.31\linewidth]{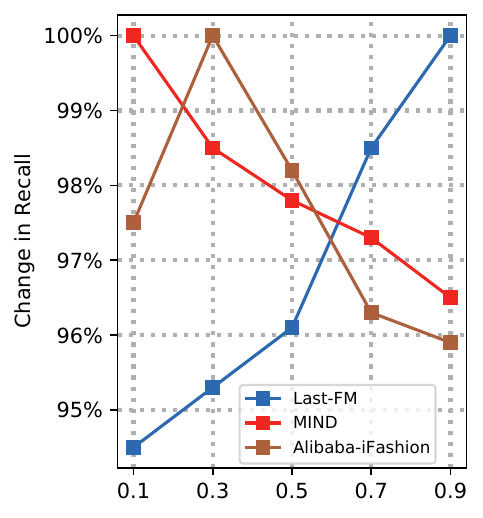}}
% \vspace{-0.2in}
\caption{Hyperparameter Study of \model.}
\label{fig:hp}
% \vspace{-0.15in}
\end{figure}

% In this part, we study how sensitive is \model~towards change in different key hyperparameters. We select the masking size $k_m$, the keep ratio for CL graph augmentation $\rho$ and the temperature for CL $\tau$. Note that for simplicity, we tune the CL keep ratio for the KG and u-i graph $\rho_k$ and $\rho_u$ simultaneously as $\rho$. From the reported results in Figure~\ref{fig:hp}, we can conclude that the best hyperparameter settings are decided by the data characteristics. Specifically, for the masking size $k_m$, 512 is the best setting for MIND and Alibaba-iFashion, and 256 for Last-FM. For the CL keep ratio, the best value is 0.5 for Last-FM and Alibaba-iFashion. And for the temperature, the best values have a large difference, as 0.1 for MIND, 0.3 for Alibaba-iFashion and 0.9 for Last-FM. The potential reason is that the denser the dataset, the more likely that the random negative samples are false-negative. Therefore, we suggest higher temperature for dense datasets (\eg Last-FM), and lower for sparse datasets (\eg MIND and Alibaba-iFashion). For masking size and CL keep ratio, we suggest tuning in $[128,512]$ and $[0.4,0.6]$.

In this study, we investigate the sensitivity of \model~to changes in key hyperparameters, including the masking size $k_m$, the keep ratio for CL graph augmentation $\rho$, and the temperature for CL $\tau$. Our analysis reveal that the optimal hyperparameter settings are highly dependent on the characteristics of the underlying data. Specifically, we found that a masking size of 512 is ideal for MIND and Alibaba-iFashion, while 256 is optimal for Last-FM. Moreover, a CL keep ratio of 0.5 is the best choice for Last-FM and Alibaba-iFashion, while a temperature of 0.1 is recommended for MIND, 0.3 for Alibaba-iFashion, and 0.9 for Last-FM. We hypothesize that this difference in optimal temperature is due to the sparsity of the datasets, with denser datasets requiring higher temperatures to avoid false-negative samples. We suggest tuning the masking size and CL keep ratio in the ranges of $[128,512]$ and $[0.4,0.6]$, respectively, as a good starting point for tuning hyperparameters in other datasets. Although \model~is relatively robust to small changes in hyperparameters, selecting the optimal settings is still critical for achieving the best performance.

\subsection{Explainability Study}
\label{sec:a:case_study}
% In this part, we investigate the explainability of recommendation results generated by \model\ by case studies on knowledge rationalization. Here, we group news items on the MIND dataset by their preset categories, and obtain the learned knowledge rationale scores for triplets connected to items within the category. From an interpretable perspective, we calculate the average of rationale scores by triplet sets of the same relation $r$, and present the cases in Table~\ref{tab:casestudy}. We collect cases from five popular news categories as their original name in Microsoft News, separately \textit{sports, newspolitics, travel, finance} and \textit{tv-celebrity}. For each category, we present two of the relations with highest average global rationale scores of their belonged triplets. From the cases, we can observe that \model~can clearly capture the impact of user interests on the KG as rationales. 

In this section, we examine the interpretability of \model's recommendation results through case studies on knowledge rationalization. Specifically, we group news items in the MIND dataset by their preset categories and obtain the learned knowledge rationale scores for triplets connected to items within the same category. To provide an interpretable perspective, we calculate the average of rationale scores by triplet sets of the same relation $r$ and present the cases in Table~\ref{tab:casestudy}. We select cases from five popular news categories, namely \textit{sports, newspolitics, travel, finance}, and \textit{tv-celebrity}. For each category, we showcase two of the relations with the highest average global rationale scores of their associated triplets. Our analysis reveals that \model~is capable of effectively capturing the impact of user interests on the KG as rationales.

% For example, when it comes to sports news, users pay more attention to league categories and specific teams, thus these two types of relations in the KG are rationalized by the labels of user preferences. The case on \textit{newspolitics} is another representative example that illustrates user preferences for political news tend to have a strong partisan orientation, and they also care about the position of political figures. Overall, the cases justify the explainability of our \model~design. By explicitly modeling the hierarchical rationality in the KG, \model~can further distinguish task rationales that are reflective of user interests. Additionally, the masking-reconstructing mechanism and cross-view rationale contrastive learning further highlight and reinforce the rationale connections.
% \vspace{-0.1in}

For instance, in the realm of sports news, users tend to focus on league categories and specific teams, and as such, these two types of relations in the knowledge graph are rationalized by the labels of user preferences. Similarly, the case of \textit{newspolitics} demonstrates that users' political news preferences often have a strong partisan orientation, and they are also concerned with the positions of political figures. These examples highlight the explainability of our \model~design. By explicitly modeling the hierarchical rationality in the knowledge graph, our approach can differentiate task rationales that reflect user interests. Moreover, the masking-reconstructing mechanism and cross-view rationale contrastive learning techniques help to emphasize and strengthen the rationale connections. This not only enhances the model's interpretability but also improves its performance by leveraging user preferences to make more accurate predictions. In summary, the rationalized knowledge graph and the \model~architecture provide a robust framework for personalized recommendation that considers user preferences and interests in a structured and transparent manner.

\end{document}